\documentclass[11pt,amstex,aps,amsfonts,prsty,preprint]{article}
\usepackage{graphicx}
\usepackage{dcolumn}
\usepackage{bm}
\usepackage{amssymb}
\usepackage{amsmath}
\usepackage{a4wide}
\usepackage{citesort}
\newcommand{\be}{\begin{equation}}
\newcommand{\ee}{\end{equation}}
\newcommand{\ud}{\mathrm{d}}

\begin{document}

\begin{titlepage}

\begin{center}
\vspace{0.3in}

{\LARGE \bf Exactly solvable model of reactions on a random catalytic chain\footnote{This work is 
dedicated to the memory of our collaborator and colleague Professor 
Alexander A. Ovchinnikov, deceased on March 5, 2003.}.}

\vspace{0.45in}

{\Large  G.Oshanin$^1$, O.B\'enichou$^2$ and A.Blumen$^3$}

\vspace{0.45in}

{ \sl $^1$ Laboratoire de Physique Th{\'e}orique des Liquides, \\
Universit{\'e} Paris 6, 4 Place Jussieu, 75252 Paris, France}

\vspace{0.15in}

{\sl $^2$ Laboratoire de Physique de la Mati{\`e}re Condens{\'e}e, \\
Coll{\`e}ge de France, 11 Place M.Berthelot, 75252 Paris Cedex 05, France
}

\vspace{0.15in}

{ \sl $^3$ Theoretische Polymerphysik,\\ Universit{\"a}t Freiburg, Hermann-Herder-Str. 3,\\
  D-79104 Freiburg, Germany}

\vspace{0.2in}

\begin{abstract}
In this paper we study a catalytically-activated $A + A \to 0$
reaction taking place on a one-dimensional regular lattice which 
is brought in contact
with  a reservoir of 
$A$ particles. 
The $A$ particles have a hard-core and
undergo continuous exchanges with the reservoir,
adsorbing onto the lattice or desorbing back to the reservoir. 
Some lattice sites possess special, catalytic properties, 
which induce an immediate reaction between 
two neighboring $A$ particles as soon as 
at least one of them lands onto a catalytic site. 
We consider three situations for the spatial placement of the catalytic sites:
regular, $annealed$ random and $quenched$ random. For all these cases
we derive 
$exact$ results for the 
partition function, and the
disorder-averaged pressure per lattice site. 
We also present exact asymptotic results 
for the particles' mean density and the
system's
compressibility.
The model studied here 
furnishes another example
of a 1D Ising-type system 
with random multisite
interactions 
which admits an exact solution.
\end{abstract}

\vspace{0.6in}
\end{center}

\end{titlepage}


\section{Introduction.}

Reactions involving particles 
which may recombine only
when some third substance - the catalytic substrate - is present  \cite{1a,7a}, but 
otherwise remain chemically inactive, are ubiquitous in nature
and also widely used in a variety of technological and industrial processes.
Within the past two decades much effort has been put  
in understanding the peculiarities of 
such catalytically-activated reactions (CARs).
In particular, 
considerable theoretical knowledge was
gained from an extensive study of
 a particular
reaction scheme -  the CO-oxidation in the presence of 
metal surfaces with  catalytic properties 
\cite{3a} (see also Ref.\cite{dic} for a recent review).
Remarkably,
Refs.\cite{3a} have substantiated 
the emergence of an 
essentially different behavior as compared to the
predictions of the classical, formal-kinetics  scheme
and have shown that under certain conditions
such
collective phenomena 
as phase transitions
or the formation of
 bifurcation patterns may take place \cite{3a}.
Prior to these works on catalytic systems,  
anomalous behavior was
amply demonstrated in other 
schemes \cite{4a,5a,6a}, 
involving reactions on contact between two particles 
 at any point of the reaction volume 
(i.e., "completely" catalytic sysems).
It
was realized \cite{4a,5a,6a} 
that the departure from the text-book, formal-kinetic predictions is
due to 
 many-particle
effects, associated with fluctuations in the spatial
 distribution of the reacting species.  
This suggests that, 
similar to such
 "completely" catalytic
reaction 
schemes,
 the behavior of the 
CARs
 may be influenced
 by many-particle
effects.  

Apart from 
many-particle effects, the
behavior of the CARs 
might be affected
by the very 
structure of the catalytic substrate, which often cannot
be considered as being a
well-defined geometrical object, 
but  represents rather an assembly
of  mobile or localized
 catalytic
sites or islands, whose spatial distribution 
is complex \cite{1a}.
  Metallic catalysts, for instance, 
are often disordered
compact aggregates, the building blocks 
of which are imperfect crystallites
with broken faces, kinks and steps. 
Usually only the steps are 
active in promoting
the reaction.
In porous materials with convoluted surfaces,
such as, e.g.,
silica, alumina or carbons, the effective catalytic
 substrate is also only a portion of the total surface area 
because of the selective participation of different sites in reaction. 
Finally, 
for  liquid-phase 
CARs the 
catalyst can consist of active  groups attached 
to polymer chains in solution.

Such complex morphologies render the 
theoretical analysis difficult. 
There are only a 
few available 
studies which 
concern 
disordered substrates such as found in the
CO-oxidation 
scheme;
here the disorder is believed
to affect
mainly 
the particles' 
adsorption and desorption 
\cite{frach,head,alb0,alb1,red1,alb2,cor,hoe,hua}.
On the other hand, 
for the situations in 
which the spatial distribution 
of 
the catalyst is random, 
only empirical approaches have been used, 
based mostly 
on  heuristic concepts of
effective reaction order
or on phenomenological 
generalizations of the formal-kinetic "law of mass action" 
(see, e.g., Refs.\cite{1a} and \cite{7a} for more details).  
The important outcome of 
such descriptions
is that they provide an evidence 
 of the existing correlations
between the
morphology of the chemically 
reactive environment
and reaction kinetic and steady-state properties. 
On the other hand,
their shortcoming is that they do not explain 
the mechanisms underlying the
anomalous kinetic and stationary behavior. 
In this regard, exact 
analytical solutions of even somewhat 
idealized or simplified 
models, are already 
highly desirable since such studies may
  provide an understanding 
of the  effects of different factors on 
the properties of the CARs.

In this paper we study 
the catalytically-activated annihilation
$A + A \to 0$ reaction 
in  a simple, one-dimensional
model 
with different (regular or random)
distributions 
of the catalysts, appropriate to the 
just mentioned situations 
with the catalytically-activated reactions
assisted by the active groups attached to polymer chains.
More specifically, we consider
here the $A + A \to 0$ reaction 
on a one-dimensional regular lattice which is brought in contact with a reservoir of $A$ partilces.
Some portion of the lattice sites (marked by crosses 
in Fig.1) possesses special "catalytic" properties such that
they induce an immediate
 reaction $A + A \to 0$, 
when 
at least one of two neighboring adsorbed $A$ particles sits on a
catalytic site. In this case these two particles react and
instantaneously leave the chain. 

In regard to the distribution
of the catalytic sites, we focus here 
on three different
situations. First, we consider the case when the catalytic sites are placed periodically, forming
a regular sublattice. Next, we turn to the disordered case. We analyse first 
the case of annealed disorder and 
furnish an exact solution. Lastly, we
analyse the behavior in the  most complex 
case of quenched disorder, for which situation 
an exact solution is also derived.

We note finally 
that the 
kinetics of $A + A \to 0$ reactions involving $\it diffusive$ $A$ particles
which react upon encounters on randomly placed catalytic sites 
has been discussed already in Refs.\cite{bur,oshan1} and \cite{tox}, and a rather
surprising behavior has been found, especially in low-dimensional systems. 
A much simplier equilibrium model
of an $A + A \to 0$ reaction on a 1D chain with
randomly placed catalytic segments has been solved in
Refs.\cite{gleb1} and \cite{gleb2}, by noticing that 
here 
the average 
pressure per lattice site coincides 
exactly  
with the Lyapunov exponent of
a product of random two-by-two matrices, obtained in Ref.\cite{der3}.
Additionally, the
steady-state properties of contact 
$A + A \to 0$ 
reactions between diffusive $A$ particles, or $A + A \to 0$ 
reactions between
immobile $A$ particles reacting
via long-range reaction probabilities
in systems with external particles input have been 
presented in Refs.\cite{racz} and \cite{sander,deem}, respectively,
which analysis has 
revealed non-trivial ordering phenomena with 
anomalous input intensity dependence of the 
mean 
particle density. This anomalous behavior agrees 
with earlier experimental
observations \cite{ben}. For completely catalytic 1D 
systems, the
kinetics of 
$A + A \to 0$ reactions with immobile $A$ particles 
undergoing cooperative desorption
has been discussed in 
Refs.\cite{nic1,nic2} and \cite{nic3}.  
Exact 
solutions for $A + A \to 0$ reactions in 1D completely catalytic systems
in which
 $A$ particles perform conventional
diffusive or subdiffusive motion have been presented in 
Refs.\cite{lush} and \cite{katja}, respectively.

The paper is structured as follows: In section 2 we formulate our model 
and introduce the basic notations.
Next, in section 3 we focus on situations with regular placements 
of catalytic sites. We calculate exactly the partition function of the model, 
the pressure per site, and present as well explicit results for the particles' mean
density in the thermodynamic limit from which we determine 
the compressibility of the system. 
In section 4  we study the behavior of the system in the case of random 
$annealed$ distributions
of the catalytic sites.
We show that in this case the model reduces to a one-dimensional 
lattice gas with an 
effective three-particle repulsive interaction. We develop a combinatorial formulation
of the model which allows us to 
obtain an exact solution. We present thus an  exact expression for the
disorder-averaged pressure, as well as exact asymptotic expansions for the mean
particle density and the compressibility. 
In section 5 we turn to the very complex situation 
where the
random distribution of catalytic
sites is $quenched$. For this case, averaging the logarithm of the partition function
over the states of the quenched random variables describing the catalytic properties of
lattice sites, we find that the problem reduces  in finite lattices 
with a fixed number of catalytic
sites to an exact enumeration of all
possible interconnected clusters. The weights of 
such clusters are calculated exactly in terms of a certain
combinatorial procedure, and we  find eventually an exact expression
 for the disorder-averaged pressure in the quenched disorder case. 
Additionally, we evaluate exact asymptotic
expansions for the mean-particle density and the compressibility. We show that the
behavior of these properties differs substantially,
depending on whether the disorder is
 annealed or quenched. In particular, in systems with annealed disorder 
the mean particle density tends to unity 
when the chemical potential
$\mu \to \infty$ for any $p < 1$, where $p$ is the mean density of catalytic
sites. On the other hand, mean particle density in the quenched disorder case
tends to a finite value $(1 - p + p^2)/(1 + p^2) < 1$ as $\mu \to \infty$. As well, in
the $annealed$ disorder case the compressibility appears to be a non-monotonic
function of $p$ for any $\mu$, while in the quenched disorder case the compressibility
shows a non-monotonic behavior as a function of $p$ only for $\mu \leq \mu_{crit}
= \beta^{-1} \ln(2)$, where $\beta^{-1}$ is the temperature. 
 Lastly, in section 6 we conclude with a summary of results. Details of intermediate 
calculations are summarized in the appendices A,B and C.

\section{The model}

Consider a one-dimensional, regular lattice containing $N$ adsorption sites (Fig.1),
which is
brought in contact with a reservoir of identical, non-interacting, hard-core
$A$ particles
- a vapor phase, maintained at a constant chemical potential $\mu$.
The $A$ particles from
the vapor phase can adsorb onto vacant adsorption sites
and desorb back to the reservoir. The occupation of the "i"-th adsorption site is
described by the Boolean variable $n_i$, such that
\begin{equation}
n_i = \left\{\begin{array}{ll}
1,     \mbox{ if the "i"-th site is occupied,} \nonumber\\
0,     \mbox{ otherwise.}
\end{array}
\right.
\end{equation}
For computational convenience, we also add two special, boundary sites
 $i = 0$ and $i = N  + 1$, and stipulate that these sites 
are always unoccupied, $n_0 =
n_{N + 1} = 0$.
\begin{figure}[ht]
\begin{center}
\includegraphics*[scale=1.1]{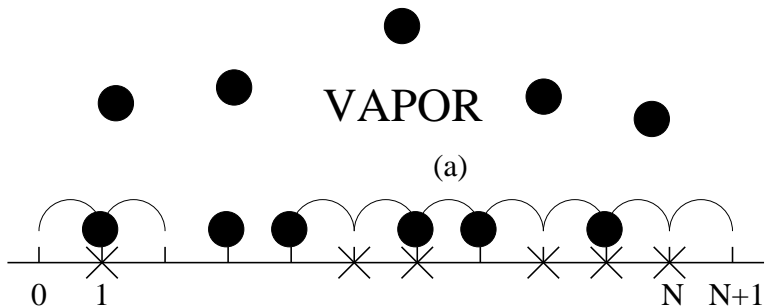}
\caption{\label{Fig1} {\small \sl One-dimensional lattice of $N$ adsorption sites in contact with a
vapor phase. The filled circles denote $A$ particles with hard-cores. The crosses denote the
adsorption sites with catalytic properties. (a) denotes a "forbidden" particle
configuration. The sites $i = 0$ and $i = N  + 1$ are always unoccupied and non-catalytic,
i.e. $n_0 =
n_{N + 1} = 0$ and $\zeta_0 =
\zeta_{N + 1} = 0$.}}
\end{center}
\end{figure}
Suppose next that some of the adsorption sites possess "catalytic" properties (crosses in Fig.1)
in the
sense that they induce an immediate
 reaction $A + A \to 0$ between neighboring $A$ particles; that
is, if at least one of two neighboring adsorbed $A$ particles sits on a
catalytic site, these two particles react and
instantaneously leave the chain (desorb
back to the reservoir). To specify the catalytic
sites, we introduce  the quenched variable
$\zeta_i$, so that
\begin{equation}
\zeta_i = \left\{\begin{array}{ll}
1,     \mbox{if the "i"-th site is catalytic, $i = 1,2, \ldots , N$, } \nonumber\\
0,     \mbox{ otherwise.}
\end{array}
\right.
\end{equation}
The 
sites at the extremities of the chain are supposed to be 
non-catalytic,
i.e. $\zeta_0 = 0$ and $\zeta_{N + 1} = 0$. As for the segment $[1,N]$, we
 will consider several possible ways of spatial placement of catalytic
sites, namely, regular and random.

For a given distribution of  catalytic sites, the
partition function $Z_N(\zeta)$ of
the system under study can be written as follows:
\begin{eqnarray}
\label{partition}
Z_N(\zeta) = \lim_{\lambda \to \infty} \sum_{\{n_i\}}
\exp\Big( \beta \mu \sum_{i=1}^N n_i -
\lambda \sum_{i = 1}^N \zeta_i \; n_i \; (n_{i-1} + n_{i+1})\Big),
\end{eqnarray}
where
the summation extends over all possible configurations
$\{n_i\}$,
$\mu$ is the chemical potential which accounts for the reservoir pressure
and for the particles' preference for
adsorption. The parameter 
$\lambda$ stands for the catalytic activity, which is here taken to
be infinitely large. 
Such a choice 
implies that the 
reaction between two neighboring 
$A$s, in 
which pair at least one of $A$s sits 
on a catalytic site, takes place instantaneously.   
Note  that here $Z_N(\zeta)$ is a functional of the
configuration $\zeta = \{\zeta_i\}$.

We stop to note that an analysis of general reaction and 
diffusion problems was previously done using effective Hamiltonians
of spin systems. Examples are, e.g.,  
the works by Doi \cite{doi}, Zeldovitch 
and Ovchinnikov \cite{zel}, 
Alcaraz et al \cite{alc} and Simon \cite{sim}. Our procedure
here is different in that we consider equilibrium systems and take the limit $\lambda \to \infty$, 
which allows us to present the basic results in closed form.

Now, taking into account that
\begin{equation}
\lim_{\lambda \to \infty} \exp\Big( - \lambda \zeta_i n_i (n_{i-1} +
n_{i+1})\Big) \equiv \Big(1 - \zeta_i n_i n_{i-1}\Big)
\Big(1 - \zeta_i n_i n_{i+1}\Big),
\end{equation}
and setting
\begin{equation}
z = \exp\Big( \beta \mu\Big),
\end{equation}
we can rewrite eq.(\ref{partition}) as:
\begin{equation}
\label{partition2}
Z_N(\zeta) = \sum_{\{n_i\}} \prod_{i=1}^N z^{ n_i}
 \Big(1 - \zeta_i n_i n_{i-1}\Big)
\Big(1 - \zeta_i n_i n_{i+1}\Big)
\end{equation}
Hence, any two neighboring sites  $i$ and $i-1$ appear to be
coupled by a factor
$(1 - n_i n_{i-1})$ when at least one of these sites is catalytic.
These coupling
factors are
depicted in Fig.1 as arcs connecting  neighboring sites
and signify that  configurations $\{n_i\}$ in which the occupation variables
$n_i$ and $n_{i-1}$ assume simultaneously the value $1$ are excluded.
We introduce now the notion of  "cluster", as being a set
of sites, all connected to each other consecutively
by arcs.
Thus, a $K$-cluster contains $K$ sites. Note that the boundary between adjacent clusters is given by a pair
of two
neighboring non-catalytic sites, i.e. by two consecutive variables $\zeta_i$ and $\zeta_{i+1}$
which are both equal to zero (see Fig.1).
Now,
the chain decomposes into disjunct clusters, and consequently,
the partition function $Z_N(\zeta)$ factorizes into independent terms, such that
each one depicts its corresponding cluster.

It may be also instructive to rewrite   eq.(\ref{partition}) in terms of the
Ising-type spin variables $\sigma_i = (2 n_i - 1)$, such that $\sigma_i = \pm 1$. In terms
of these variables,  $Z_N(\zeta)$ of eq.(\ref{partition}) reads:
\begin{equation}
Z_N(\zeta) = \lim_{\lambda \to \infty} \exp\Big( (\beta \mu - \lambda p) \frac{N}{2} \Big)  \sum_{\{\sigma_i\}}
\exp\Big(\sum_{i=1}^N \mu_i \sigma_i + \sum_{i=1}^N J_i \sigma_i \sigma_{i+1}\Big)
\end{equation}
where $p$ denotes the mean density of catalytic sites, $p = N^{-1} \sum_{i}^N
\zeta_i$, while
\begin{equation}
\mu_i = \frac{\beta \mu}{2} - \frac{\lambda}{4} (2 \zeta_i + \zeta_{i -1} + \zeta_{i+1}),
\;\;\; \text{and} \;\;\; J_i = - \frac{\lambda}{4} (\zeta_i + \zeta_{i+1}).
\end{equation}
Consequently, the model under study can be also thought of as a version
of a one-dimensional Ising-type model with site-dependent magnetic fields
and site-dependent couplings (see, e.g.,  some seminal works \cite{der1,der2,der3}, as well
Refs.\cite{ranis} and \cite{vik} for a
recent review). Note, however, that in our case both the fields and the couplings are
non-local, and that the local energy $\epsilon_i$, 
$\epsilon_i = - \mu_i \sigma_i - J_i \sigma_i \sigma_{i+1}$,
can assume several different values, depending on the occupation variables of the neighboring sites
and on their catalytic properties.

We close this section by mentioning the results of the conventional
mean-field approach, which depicts the
evolution of our system
\cite{1a,7a}. Discarding correlations
 between the occupation of 
neighboring sites, i.e. setting $n_i = \overline{n}$ , one writes the following balance equation
\begin{equation}
\label{meanfield}
\dot{\overline{n}} = - p {\cal  K} \overline{n}^2 + g_{ads} \left(1 - \overline{n}\right) - g_{des} \overline{n},
\end{equation}
where the overdot denotes the time derivative, ${\cal K}$ stands for the 
elementary reaction
constant, $p$ is the mean density
of catalytic sites (introduced 
to account for
the reduction in the reaction 
rate due to the partial chain coverage by catalytic sites),
while the second and the third terms on the rhs of eq.(\ref{meanfield}) correspond
to the usual Langmuir adsorption/desorption events; here $g_{ads}$ and 
$g_{des}$ are the adsorption and desorption rates, respectively, $g_{ads}/g_{des} = z$. 

Equation (\ref{meanfield}) has the following equilibrium 
solution
\begin{equation}
\label{no}
\overline{n} = \frac{\left(g_{ads} +
 g_{des}\right)}{2 p {\cal K}} \left\{\sqrt{\frac{4 p {\cal  K} g_{ads}}{\left(g_{ads}+g_{des}\right)^2}+1} - 
1\right\}.
\end{equation}
In the limit ${\cal K} \to \infty$, 
(which is the limit of interest in the 
present paper),  and with $p$ and $g_{ads}$ kept finite one 
finds that $\overline{n} \to 0$; from eq.(\ref{no}) $\overline{n}$ vanishes as
$\overline{n} \sim \sqrt{g_{ads}/p {\cal K}}$. 
In the following we proceed to 
show that the actual behavior of the mean density of the $A$ particles turns out to be
very different from eq.(\ref{no}); this is 
due to the emerging correlations between the $A$ particles. 
Note also that in the limit ${\cal K} \equiv 0$ (suppressed reaction), 
one recovers from eq.(\ref{no}) the classic Langmuir 
result $\overline{n} = z/(1+z)$ \cite{1a,7a}. 

\section{Regular placement of catalytic sites.}

To fix the ideas,
consider first a situation in which  the catalytic sites are placed periodically, with
period $L$, so that $\zeta_i$ obeys:
\begin{equation}
\label{ze}
\zeta_i = \delta(i,n L + 1), \; \; \;\text{with} \;\;\;  n = 0,1, \ldots, \Big[\frac{N - 1}{L} \Big],
\end{equation}
where $[x]$ denotes the integer part of
the number $x$, and
$\delta(k,m)$ is the Kroneker-delta symbol,
\begin{equation}
\label{delta}
\delta(k,m) = \frac{1}{2 \pi i} \oint_{{\cal C}} \frac{\ud \tau}{\tau^{1+k-m}}
= \left\{\begin{array}{ll}
1,     \mbox{if $k = m$,} \nonumber\\
0,     \mbox{ otherwise,}
\end{array}
\right.
\end{equation}
where ${\cal C}$ stands for any closed contour which encircles the origin counterclockwise while 
$(k,m)\in\mathbb{Z}^2$.

\begin{figure}[ht]
\begin{center}
\includegraphics*[scale=1.1]{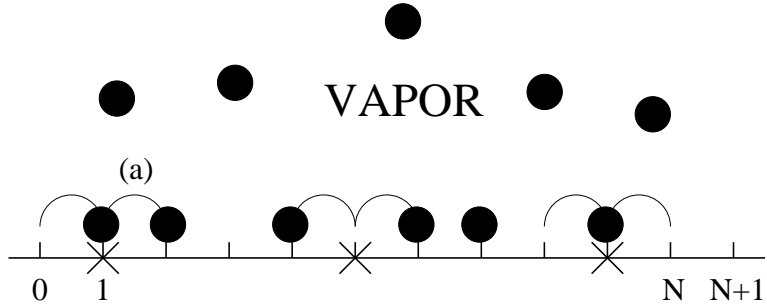}
\caption{\label{Fig2} {\small \sl Periodic placement of the catalytic sites with the
period $L = 4$. (a) denotes a forbidden particle configuration.}}
\end{center}
\end{figure}
We have now to distinguish between two situations: namely, when $L \geq 3$ and when $L
= 1$ or $L = 2$. In the former case, evidently, the factors $(1 - \zeta_i n_i n_{i \pm 1})$
in eq.(\ref{partition2})
are non-overlapping (see Fig.2); then the partition function decomposes into
elementary three-clusters centered around each catalytic site and (possibly) into
uncoupled, "free" sites,
i.e. sites unaffected by any of the factors $(1 - \zeta_i n_i n_{i \pm 1})$.
On the other hand,
in the cases $L = 1$ and $L = 2$
we deal with totally interconnected clusters, spanning the
entire chain, as one can deduce from Fig.3. In fact, the role of $L = 1$ and $L = 2$ is,
chemically speaking, identical.
\begin{figure}[ht]
\begin{center}
\includegraphics*[scale=1.1]{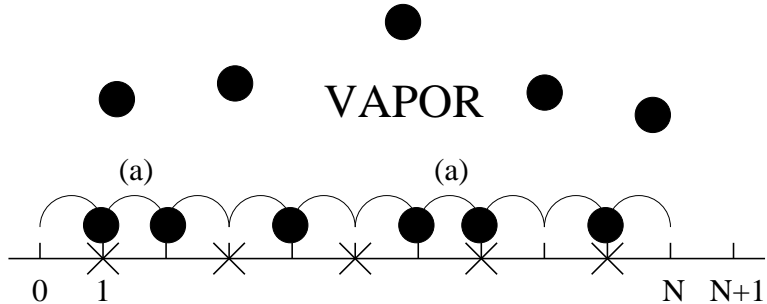}
\caption{\label{Fig3} {\small \sl Periodic placement of the catalytic sites with the
period $L = 2$. In this case all sites are coupled by factors $(1- n_i n_{i-1})$ and
hence the occupation variables of any two neighboring sites can not assume
 the value $1$ simultaneously; (a) denotes such  "forbidden" particle
configurations.}}
\end{center}
\end{figure}

\subsection{Periodic case with $L \geq 3$.}

In the case $L \geq 3$ the partition function in
eq.(\ref{partition2})
decomposes into the product
\begin{equation}
\label{deco}
Z_N(\zeta) = Z_N^{(reg)}(L) =
Z_3^{{\cal N}_3} \;  Z_2^{{\cal N}_2} \; Z_1^{{\cal N}_1},
\end{equation}
where the superscript "(reg)" signifies that we deal with the regular case,
and where $Z_K$, $K = 1$, $K = 2$ and $K = 3$,
are the partition functions of one-, two-  and three-clusters,
while ${\cal N}_K$, $K = 1,2,3$,
stands for the numbers of such clusters in the
$N$-chain, respectively. Now:
\begin{equation}
{\cal N}_3 = \Big[\frac{N - 1}{L}\Big] - \delta\Big(\frac{N - 1}{L},\Big[\frac{N - 1}{L}\Big]\Big); \;\;\;
{\cal N}_2 = 1 + \delta\Big(\frac{N - 1}{L},\Big[\frac{N - 1}{L}\Big]\Big),
\end{equation}
and
\begin{equation}
\label{t}
{\cal N}_1 = N  -  2 +  \delta\Big(\frac{N - 1}{L},\Big[\frac{N - 1}{L}\Big]\Big) - 3 \Big[\frac{N - 1}{L}\Big],
\end{equation}
where we have used the evident "conservation" law
\begin{equation}
3 {\cal N}_3 + 2 {\cal N}_2 + {\cal N}_1 = N.
\end{equation}
Note, however, that the number of the two-clusters is not extensive,
i.e., it does not grow with $N$: Such
clusters can be  present only on the  boundaries, i.e.,
 ${\cal N}_2 = 2$ when $(N - 1)/L$ is an integer and
${\cal N}_2 = 1$ otherwise. Moreover,
\begin{equation}
\label{rec1}
Z_1 = \sum_{\{n_1 = 0,1\}} z^{n_1} = (1 + z); \;\;\; 
Z_2 =
\sum_{\{n_1,n_2 = 0,1\}} z^{n_1 + n_2} (1 - n_1 n_2)
=  (1 + 2 z),
\end{equation}
and
\begin{equation}
\label{rec3}
Z_3 = \sum_{\{n_1,n_2,n_3 = 0,1\}} z^{n_1 + n_2 + n_3} (1 - n_1 n_2) (1 - n_2 n_3)
=  (1 + 3 z + z^2).
\end{equation}
Consequently,  in the case of a regular, periodic placement of  catalytic sites with
period $L$, $L \geq 3$, we can calculate  
the pressure $P^{(reg)}(L)$ per site
from the relation:
\begin{equation}
\label{free}
 \beta P^{(reg)}(L) =  \lim_{N \to \infty} \frac{\ln Z_N^{(reg)}(L)}{ N},
\end{equation}
which yields:
\begin{eqnarray}
\label{free-reg}
 \beta P^{(reg)}(L) &=&   \lim_{N \to \infty} \Big( \frac{{\cal N}_3}{N} \ln Z_3
- \frac{{\cal N}_2}{N} \ln Z_2 - \frac{{\cal N}_1}{N} \ln Z_1\Big)
= \nonumber\\
&=& p \; \ln(1 + 3 z + z^2) + (1 - 3 p) \; \ln(1 + z),
\end{eqnarray}
where $p = 1/L$ is the density of  catalytic sites.

The averaged density of adsorbed particles obeys
\begin{equation}
\label{isotherm}
n^{(reg)}(L) = \frac{1}{N} \overline{\sum_{i =1}^N n_i}
= \frac{z}{N} \frac{d}{d z} \ln Z_N^{(reg)}(L),
\end{equation}
and hence, in the limit $N \to \infty$ it follows that
\begin{equation}
\label{isotherm-reg}
n^{(reg)}(L) = (1 - 3 p) \frac{z}{1 + z} + p \frac{3 z + 2 z^2}{1 + 3 z + z^2}.
\end{equation}
We stop to note that in the limit $p = 0$ (i.e. that of a 
completely non-catalytic chain)
eq.(\ref{isotherm-reg}) reduces to the standard Langmuir result. On the other hand,
when $p = 1/3$ the number of free-sites in the periodic chain vanishes and the Langmuir
 contribution gets equal to zero; in this case the chain is composed totally of
three-clusters. In this case, for very large $z$ the occupation of lattice 
sites by adsorbed molecules equals
$2/3$, as it should.

Finally, we have that the compressibility $k_T$, defined as
\begin{equation}
\beta^{-1} k_T = \frac{1}{n^2} \frac{\partial n}{\partial \mu} = \frac{z}{n^2} \frac{\partial n}{\partial z}
\end{equation}
obeys
\begin{equation}
\label{comp}
\beta^{-1} k_T^{(reg)}(L) = \frac{1 + 6 z + 11 z^2 + 6 z^3 + z^4 - p z (8 + 8 z + 19 z^2)}{z (1 + 3 z + z^2 - 4 p z - p z^2)^2}.
\end{equation}
For $p = 0$ the last
equation reduces to the standard 
Langmuir result, $k_T^{(lan)} = \beta/z$.

\subsection{Periodic cases with $L = 1$ or $L = 2$.}

We note first that since $(1 - n_i n_{i+1})^2 = (1 - n_i n_{i+1})$, we evidently have
that
\begin{equation}
\label{partition3}
Z_N \equiv Z_N^{(reg)}(L = 1) = Z_N^{(reg)}(L = 2) =
\left. \sum_{\{n_i\}, i = 1, \ldots ,N}  z^{\sum_{i = 1}^N n_i} \prod_{i=1}^{N}
 \Big(1 - n_i n_{i+1}\Big)\right|_{ n_{N+1}  = 0}.
\end{equation}
We hence conclude that
the $L = 1$ and the $L = 2$ periodic systems are chemically equivalent.

Now, we focus on the derivation of the form of $Z_N$  according to eq.(\ref{partition3}).
To do this, we proceed as
follows: performing first the averaging over the states of the variable $n_N$, we have
that $Z_N$ can be written as
\begin{equation}
\label{part}
Z_N = Z_{N - 1} + z
\sum_{\{n_i\}, i = 1, \ldots, N-1} z^{\sum_{i=1}^{N-1} n_i}
\prod_{i=1}^{N - 2} \Big(1 - n_i n_{i+1}\Big) \Big(1 - n_{N - 1}\Big),
\end{equation}
where $Z_{N-1}$ obeys
\begin{equation}
\label{partition5}
Z_{N-1}  =
\left. \sum_{\{n_i\}, i = 1, \ldots ,N - 1}  z^{\sum_{i = 1}^{N - 1} n_i} \prod_{i=1}^{N - 1}
 \Big(1 - n_i n_{i+1}\Big)\right|_{n_{N}  = 0},
\end{equation}
and is hence
the partition function of
a chain consisting of $N-1$ sites. We
notice next that the second term on the rhs of eq.(\ref{part})
vanishes when $n_{N -
1} = 1$;  thus it can be written as
\begin{eqnarray}
\label{part2}
&&\sum_{\{n_i\}, i = 1, \ldots, N-1}  z^{ \sum_{i=1}^{N -1} n_i}
\prod_{i=1}^{N - 2} \Big(1 - n_i n_{i+1}\Big) \Big(1 - n_{N - 1}\Big) = \nonumber\\
&=& \left. \sum_{\{n_i\}, i = 1, \ldots, N-2}  z^{ \sum_{i=1}^{N -2} n_i}
\prod_{i=1}^{N - 2} \Big(1 - n_i n_{i+1}\Big)\right|_{ n_{N -1}  = 0},
\end{eqnarray}
where the expression on the right-hand-side  is, evidently, the partition function of a chain containing $N - 2$
sites. Consequently, the partition function of an $N$-site chain obeys the following recursion:
\begin{equation}
\label{recursion}
Z_N =  Z_{N - 1} +
z Z_{N - 2}
\end{equation}
whose first three terms  are given by eqs.(\ref{rec1}) and (\ref{rec3}).

Next, to determine $Z_N$ explicitly for arbitrary $N$ we introduce the
following generating function
\begin{equation}
\label{gen}
{\cal H}(\tau) = \sum_{N = 1}^{\infty} Z_N \tau^N.
\end{equation}
Then, multiplying both sides of eq.(\ref{recursion}) by $\tau^{N -2}$ and performing the summation,
we  obtain the following explicit result for ${\cal H}(\tau)$:
\begin{eqnarray}
\label{gene}
{\cal H}(\tau) &=& \Big(z \tau^2 +(1+z) \tau\Big) \Big(1 - \tau - z \tau^2\Big)^{-1} = \nonumber\\
&=& \frac{\Big(z \tau^2 + (1 + z) \tau\Big)}{\sqrt{1 + 4 z}} \Big(\frac{1}{\tau + \tau_1}
+ \frac{1}{\tau_2 - \tau} \Big),
\end{eqnarray}
where $\tau_1$ and $\tau_2$ are given by
\begin{equation}
\label{tau1}
\tau_{1} = \frac{1}{2 z} \Big(\sqrt{1 + 4 z} + 1\Big)
\end{equation}
and
\begin{equation}
\label{tau2}
\tau_{2} = \frac{1}{2 z} \Big(\sqrt{1 + 4 z} - 1\Big).
\end{equation}
Next, expanding the terms in the second line on the rhs of eq.(\ref{gene}) in a Taylor
series in powers of $\tau$ and gathering
terms entering with the same power, we find that  $Z_N$ obeys
\begin{equation}
\label{explicit}
Z_N = \frac{1 + 2 z + \sqrt{1 + 4 z}}{2  \sqrt{1 + 4 z} \;  \tau_{2}^N} \; {\cal L}_N,
\end{equation}
where
\begin{equation}
\label{L}
{\cal L}_N = \Big(1 - (-1)^N
\frac{(1 + 2 z - \sqrt{1 + 4 z})}{(1 + 2 z + \sqrt{1 + 4 z})} \Big(\frac{\tau_2}{\tau_1}\Big)^N \Big)
\end{equation}
We recall that the $Z_N$ are, of course, 
polynomial functions of the activity $z$;  expanding  
the rhs of eq.(\ref{explicit}) in powers of $z$, we get
\begin{equation}
\label{pol}
Z_N = \sum_{l =0}^{[(N+1)/2]} {N-l+1
\choose
l}    z^l,
\end{equation}
where ${N
\choose
l}$ denote the binomial coefficients,
\begin{equation}
\label{binomial}
{N
\choose
l}
= \left\{\begin{array}{ll}
\displaystyle N!/l!(N-l)!,     \mbox{for $N \geq l$,} \nonumber\\
0,     \mbox{ otherwise.}
\end{array}
\right.
\end{equation}
For large values of $z$, it might be more convenient to use
another representation; subtracting in eq.(\ref{pol}) $z^{(N+1)/2}$ and summing up
the remaining terms, we find that $Z_N$ are given by
\begin{equation}
Z_N =  z^{(N+1)/2} F_{N + 2}(1/\sqrt{z}),
\end{equation}
where $ F_{n}(x)$ are the Fibonacci polynomials \cite{fibo}, defined explicitly by
\begin{equation}
\label{fibo}
F_n(x) = \sum_{l =0}^{[(n-1)/2]} {n-l-1
\choose
l}    x^{n-2l-1}.
\end{equation}

Finally, noticing that for $z < \infty$ one has $(\tau_2/\tau_1) < 1$ and hence, that $(\tau_2/\tau_1)^N$ vanishes 
exponentially with $N$ as $N \to \infty$,
we find  in
the cases when $L = 1$ and $L =2$ that the pressure per site obeys, using eq.(\ref{free}):
\begin{equation}
\label{i}
\beta P^{(reg)}(L = 1  \, {\rm or} \,  2) =  \ln\Big(\frac{\sqrt{1 + 4 z} + 1}{2}\Big),
\end{equation}
while the average density in an infinite chain is, using eq.(\ref{isotherm}):
\begin{equation}
\label{"}
n^{(reg)}(L = 1 \, {\rm or} \, 2) = 1 - \frac{2 z}{1 + 4 z - \sqrt{1 + 4 z}}.
\end{equation}
In the limit $z = \infty$, the roots $\tau_1$ and $\tau_2$ are equal, $\tau_1 = \tau_2$,
this
signals the emergence of
long-range order. Actually, in this case $n^{(reg)}(L = 1  \, {\rm or} \, 2) = 1/2$ and
the particles' distribution on the lattice is periodic.

Finally, to close this section, we derive the 
compressibility 
of the particle phase. 
In the thermodynamic limit, for 
regular placement of the catalytic
sites with 
the period $L = 1$ or $L = 2$ it obeys:
\begin{equation}
\label{comp1}
\beta^{-1} k_T^{(reg)}(L =1  \, {\rm or} \,  2)  = \frac{2 z}{\sqrt{1 + 4 z} (1 + 2 z - \sqrt{1 + 4 z})}.
\end{equation}
Note that contrary to the expression in 
eq.(\ref{comp}) which holds for $L \geq 3$, here $k_T^{(reg)}(L = 1  \, {\rm or} \, 2)$ is a non-analytic function
of the activity $z$ when $z \to \infty$.

\section{Random placement of catalytic sites. Annealed disorder.}

Consider next the situation with $\it annealed$ disorder, which is realized, for instance,
when the catalytic property may move (diffuse) very quickly. 
In this case, the disorder-average pressure $P^{(ann)}(p)$
per site and the average density are given by, respectively,
\begin{equation}
\label{ann}
P^{(ann)}(p) = \lim_{N \to \infty} \frac{\ln \left<Z_N(\zeta)\right>}{\beta N},
\end{equation}
and
\begin{equation}
\label{density}
n^{(ann)}(p) = \frac{1}{N} \overline{\sum_{i =1}^N n_i}
=  \beta z \frac{d}{d z} P^{(ann)}(p),
\end{equation}
where $Z_N(\zeta)$ is again the partition
function of eq.(\ref{partition}).

We can now average directly the partition function in eq.(\ref{partition}) over the
placement of the sites with
catalytic property:
\begin{eqnarray}
 \left<Z_N(\zeta)\right> &=& \lim_{\lambda \to \infty} \sum_{\{n_i\}}
\exp\Big( \beta \mu \sum_{i=1}^N n_i\Big) \left< \exp\Big(
- \lambda \sum_{i = 1}^N \zeta_i n_i (n_{i-1} + n_{i+1})\Big)\right> = \nonumber\\
&=&  \lim_{\lambda \to \infty} \sum_{\{n_i\}}
\exp\Big( \beta \mu \sum_{i=1}^N n_i\Big) \prod_{i = 1}^{N} \left<\exp\Big(- \lambda
\zeta_i n_i (n_{i-1} + n_{i+1})\Big)\right> = \nonumber\\
&=& \sum_{\{n_i\}} z^{\sum_{i=1}^N n_i}  \lim_{\lambda \to \infty} \prod_{i = 1}^{N}
\Big( p \exp\Big(- \lambda  n_i (n_{i-1} + n_{i+1})\Big) + 1 - p\Big).
\end{eqnarray}
Noticing next that
\begin{equation}
\lim_{\lambda \to \infty} \Big(p \exp\Big(- \lambda  n_i (n_{i-1}
+ n_{i+1})\Big) + 1 - p\Big) = \left\{\begin{array}{ll}
1 - p,     \mbox{ if $ n_i (n_{i-1} + n_{i+1}) > 0$,} \nonumber\\
1,     \mbox{if   $ n_i (n_{i-1} + n_{i+1}) = 0$,}
\end{array} \right.
\end{equation}
and hence, that
\begin{equation}
\lim_{\lambda \to \infty} \Big(p \exp\Big(- \lambda  n_i (n_{i-1}
+ n_{i+1})\Big) + 1 - p\Big) = (1 - p)^{\displaystyle n_i (1 - (1 - n_{i-1})(1 - n_{i+1}))},
\end{equation}
we find that
the averaged partition function in eq.(\ref{partition}) attains the form
\begin{equation}
\label{zu}
 \left<Z_N(\zeta)\right> = \sum_{\{n_i\}}
z^{\sum_{i=1}^N n_i} (1 - p)^{\sum_{i=1}^N \Psi_i},
\end{equation}
where $\Psi_i$ is a Boolean variable:
\begin{equation}
\label{psipsi}
\Psi_i = (n_i n_{i+1} + n_i n_{i-1} -
n_{i-1} n_i n_{i+1}) = \left\{\begin{array}{ll}
1,     \mbox{ if $ n_i (n_{i-1} + n_{i+1}) > 0$,} \nonumber\\
0,     \mbox{if   $ n_i (n_{i-1} + n_{i+1}) = 0$.}
\end{array} \right.
\end{equation}
Evidently,  $\Psi_i$ is non-local and
depends on the environment of the "i"-th site.
As a matter of fact, the local energy $\epsilon_i$ at the "i"-th site,
$\epsilon_i = - \beta \mu n_i - \ln(1-p) (n_i n_{i+1} + n_i n_{i-1} -
n_{i-1} n_i n_{i+1})$,
assumes $8$ different values depending on $n_{i-1}$, $n_i$ and $n_{i+1}$. 
Before we proceed further, it might be also instructive to rewrite
eq.(\ref{zu}) in terms of the spin variables $\sigma_i$. Then eq.(\ref{zu}) takes
the form
\begin{eqnarray}
\label{is}
 \left<Z_N(\zeta)\right> &=& \exp\Big((\beta \mu + \frac{3}{4} \ln(1-p)) \frac{N}{2}\Big)
\sum_{\{\sigma_i\}}  \exp\Big( \beta \mu'  \sum_{i=1}^N \sigma_i + J_1
\sum_{i=1}^N \sigma_i \sigma_{i+1} - \nonumber\\
&-& J_2 \sum_{i=1}^N \sigma_{i-1} \sigma_{i+1} - J_3
\sum_{i=1}^N \sigma_{i-1} \sigma_i \sigma_{i+1} \Big),
\end{eqnarray}
where
\begin{equation}
\mu' = \mu + \frac{5}{8 \beta} \ln(1-p); \;\;\; J_1 = \frac{ \ln(1-p)}{4} < 0; \;\;\; \text{and} \;\;\; 
J_2 = J_3 = \frac{ \ln(1-p)}{8} < 0.
\end{equation}
As one may notice, the expression in eq.(\ref{is}) 
represents a combination of two
well-known Ising-type models: namely,
the first three terms in the exponent define the
antiferromagnetic ANNNI (axial next-nearest neighbor 
Ising) chain \cite{annni2}. 
In our case,
the competition ratio $k = J_2/J_1$ equals $k = 1/2$, which is, in fact,
a non-trivial special point (the so-called
multiphase point \cite{annni3}) of the ANNNI model. On the other hand, the fourth term in the exponent
corresponds to  the three-spin interaction model \cite{baxter}.
Both models have been extensively
studied within the last two decades in various contexts \cite{annni3,baxter}
and show interesting equilibrium and dynamic properties, see e.g.,  Refs.\cite{kisker,dhar}.
We are, however, unaware of
an exact solution of the one-dimensional combined model of eq.(\ref{is}). Below we
will furnish such an exact solution using a combinatorial approach.

Note now that $\Psi_i$ always equals zero
for unoccupied sites, $(n_i = 0)$,
but attains the value $\Psi_i
= 1$ only for occupied sites, $(n_i = 1)$, which have at least one (or two)
occupied neighboring sites (see Fig.4).
\begin{figure}[ht]
\begin{center}
\includegraphics*[scale=1.1]{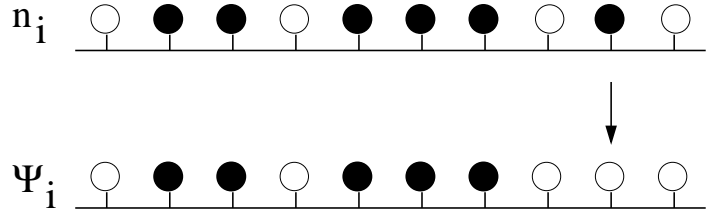}
\caption{\label{Fig4} {\small \sl Values of the variable $\Psi_i$ corresponding
to a given configuration $\{n_i\}$.}}
\end{center}
\end{figure}
Otherwise, for isolated occupied sites (elementary sequences with
$(n_{i-1}=0,n_{i}=1,n_{i+1}=0)$) the variable $\Psi_i$ equals $0$. 
Consequently, for any given realization
$\{n_i\}$, one has that
\begin{equation}
\label{dif}
\sum_{i=1}^N \Psi_i = {\cal N}_+[\{n_i\}]  - {\cal N}_{is}[\{n_i\}],
\end{equation}
where ${\cal N}_+[\{n_i\}]$ is the number of lattice
sites on which  (in a given realization
$\{n_i\}$) the occupation variable $n_i$ assumes the value $1$, while ${\cal
N}_{is}[\{n_i\}]$
is the
realization-dependent number of isolated occupied sites (elementary
cells of the form $(0,1,0)$). Hence, the partition function in eq.(\ref{zu})
can be rewritten as
\begin{equation}
\label{ro}
\left<Z_N(\zeta)\right> = \sum_{\{n_i\}}
\Big(z (1 - p)\Big)^{{\cal N}_+[\{n_i\}]}  (1 - p)^{- {\cal N}_{is}[\{n_i\}]}
\end{equation}
Next, ordering the entire set of $2^N$ different realizations $\{n_i\}$ with
respect to the total number of adsorbed particles which they contain,
i.e. ${\cal N}_+[\{n_i\}]$,
 we can recast eq.(\ref{ro}) into the form:
\begin{equation}
\label{gen}
\left<Z_N(\zeta)\right> = \sum_{{\it N}_+ =0}^{N}
\Big(z (1 - p)\Big)^{{\it N}_+} \; \sum_{m = 0}^{N - {\it N}_+ +1}
(1 - p)^{-m} M_m({\it
N}_+|N),
\end{equation}
where
$M_m({\it
N}_+|N)$ stands for the number of
realizations $\{n_i\}$
that have a fixed ${\it N}_+$
and contain $\it exactly$
$m$ elementary cells $(0,1,0)$.

\subsection{Calculation of $M_m({\it
N}_+|N)$.}

To evaluate $M_m({\it
N}_+|N)$ we now proceed as follows. Let us
consider a given realization $\{n_i\}$ with ${\it
N}_+$ filled (and ${\it
N}_-= N - {\it
N}_+$ vacant) sites and specify the lattice positions of the vacant sites by introducing a
set of intervals $\{{\it l}_j\}$, $j = 1, \ldots, {\it
N}_-+1$, such that the interval ${\it l}_1$  connects a boundary site $i = 0$ with the
first vacant site, the interval  ${\it l}_2$ connects the first vacant site with the
second one, and etc, while the last interval ${\it l}_{\it
N_- {\rm +1}}$ connects the last vacant site with the boundary site $i = N + 1$ (see Fig.5).

These intervals, which uniquely define
the positions of the
vacant sites in each given realization  $\{n_i\}$, obey the "conservation law":
\begin{equation}
\label{dio}
{\it l}_1 + {\it l}_2 + {\it l}_3 + \; \ldots \; + {\it l}_{\it
N_-{\rm +1}} = N + 1
\end{equation}
Then, since each interval containing exactly
 two lattice units corresponds to an isolated
occupied site,
$M_m({\it
N}_+|N)$
equals the number of different solutions of eq.(\ref{dio}), constrained by
the condition that in each sequence $\{{\it l}_i\}$ obeying eq.(\ref{dio})
$m$ of the ${\it
N}_{-}+1$ intervals are equal to $2$, while the rest can assume any value except 
$2$. Hence,   $M_m({\it
N}_+|N)$ can be represented as
\begin{equation}
\label{y}
M_m({\it
N}_+|N) = {{\it N}_{-}+1 \choose m} \; {\it P}_{m}({\it
N}_+|N).
\end{equation}
Here the binomial coefficient accounts for all
possible choices of $m$ intervals from  the set of ${\it
N}_{-}+1$ intervals, while ${\it P}_{m}({\it
N}_+|N)$ stands for the number of different solutions of the equation
\begin{equation}
{\it l}_1 + {\it l}_2 + {\it l}_3 + \; \ldots \; + {\it l}_{{\it
N}_{-}-m+1} = N - 2 m +1,
\end{equation}
where each of intervals ${\it l}_k$, $k = 1,2, \ldots, {\it
N}_{-}-m+1$, can assume one of the values ${\it l}_k = 1,3,4,5, \ldots $.
\begin{figure}[h]
\begin{center}
\includegraphics*[scale=1.1]{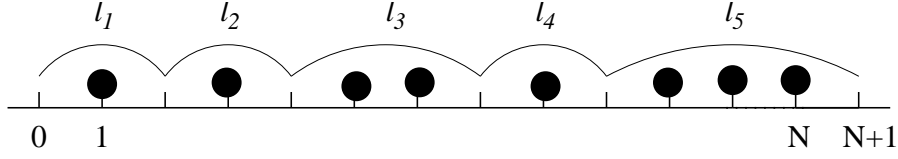}
\caption{\label{Fig5} {\small \sl Definition of a given configuration $\{n_i\}$ in terms of the
intervals $\{{\it l}_i\}$ connecting sequentially  unoccupied lattice sites.}}
\end{center}
\end{figure}
Making use of the integral representation of the lattice delta-function in
eq.(\ref{delta}),
we can write ${\it P}_{m}({\it
N}_+|N)$ as
\begin{eqnarray}
{\it P}_{m}({\it N}_+|N) = \frac{1}{2 \pi i} \oint_{{\cal C}} \frac{d \tau}{\tau} \;
\tau^{ -(N - 2 m + 1)} 
 \Big\{ \hat{\sum_{{\it l}_{\rm 1}}}  \; \ldots \; \hat{\sum_{{\it
l}_{{\it N}_{-}-m+1}}} \tau^{\displaystyle \Big({\it l}_1 + {\it l}_2 + {\it l}_3 + \; \ldots \;
 + {\it l}_{{\it
N}_{-}-m+1}\Big)}\Big\},
\end{eqnarray}
where the hat above the summation symbol signifies that summation runs
over all possible values $l_k = 1,3, \ldots, \infty$, excluding the value $l_k =
2$. Performing the summation, we find 
\begin{equation}
\label{x}
{\it P}_{m}({\it N}_+|N) =  \frac{1}{2 \pi i} \oint_{{\cal C}} \; \frac{d \tau}{\tau} \;
\tau^{\displaystyle \Big({\it N}_+ - m\Big)} \; S_{\tau}^{N - {\it N}_+ + 1 -m},
\end{equation}
where
\begin{equation}
S_{\tau} = \left(\Big(1 - \tau   \Big)^{-1} -  \tau\right).
\end{equation}

\subsection{Explicit form of $\left<Z_N(\zeta)\right>$.}

Substituting eqs.(\ref{x}) and (\ref{y}) into eq.(\ref{gen}), we 
find that the partition function obeys
\begin{equation}
\label{+}
\left<Z_N(\zeta)\right> = \frac{1}{2 \pi i} \oint_{{\cal C}} \; \frac{d \tau}{\tau} \;
S_{\tau}^{N + 1}  \;
\sum_{{\it N}_+ =0}^{N}
\Big(\frac{z (1 - p)}{\tau S_{\tau}}\Big)^{{\it N}_+} \;
\sum_{m = 0}^{{\it N}_- +1}  {{\it N}_{-}+1  \choose  m}   \;
\Big(\frac{\tau}{(1-p) S_{\tau}}\Big)^{m},
\end{equation}
which yields, upon summing over
$m$ and ${\it N}_+$, the following result
\begin{eqnarray}
\label{ZZ}
\left<Z_N(\zeta)\right> &=& \frac{2 (1 - p)}{9 (-  Q) p} \Big\{
- \frac{1 - 3 \sqrt{- Q} \cos\Big( \displaystyle \frac{1}{3}
\arccos\Big(R/\sqrt{-Q^3}\Big)\Big)}{ \displaystyle 1 + 2 \cos\Big(\frac{2}{3}
\arccos\Big(R/\sqrt{-Q^3}\Big)\Big)}  \Big( \displaystyle \frac{z (1-p)}{t_1}\Big)^{N+2} +
\nonumber\\
&+& \frac{1 + 3 \sqrt{- Q} \sin\Big( \displaystyle \frac{1}{3}
\arcsin\Big(R/\sqrt{-Q^3}\Big)\Big)}{2 \sin\Big( \displaystyle \frac{\pi}{6} + \frac{2}{3}
\arccos\Big(R/\sqrt{-Q^3}\Big)\Big) - 1}  \Big( \displaystyle \frac{z (1-p)}{
t_2}\Big)^{N+2} - \nonumber\\
&-& \frac{ \displaystyle 1 + 3 \sqrt{- Q} \sin\Big( \displaystyle \frac{1}{3}
\arcsin\Big(R/\sqrt{-Q^3}\Big)\Big)}{ \displaystyle 1 - 2 \sin\Big(\frac{\pi}{6} - \frac{2}{3}
\arccos\Big(R/\sqrt{-Q^3}\Big)\Big)}  \Big( \displaystyle \frac{z (1-p)}{
t_3}\Big)^{N+2}\Big\},
\end{eqnarray}
where $R$ and $Q$ are 
auxiliary functions, which obey
\begin{equation}
\label{Q}
Q = - \frac{1}{9} - \frac{(1-p)}{3 p}(1 + z (1-p)) \leq 0,
\end{equation}
and
\begin{equation}
\label{R}
R = \frac{1}{27} +  \frac{1}{6} \Big(  \frac{(1-p)}{p}(1 + z (1-p)) -
\frac{3 z (1-p)^2}{p} \Big).
\end{equation}
Equation (\ref{ZZ})
defines $\left<Z_N(\zeta)\right>$ for arbitrary values of 
$p$, $z$ and chain's length $N$. The derivation of eq.(\ref{ZZ}) is presented in Appendix A.

\subsection{The thermodynamic limit $N \to \infty$.}

We turn next to the thermodynamic limit aiming to calculate the disorder-averaged pressure per site
in the $annealed $ disorder case 
and the mean density of 
adsorbed particles.  In the limit $N \to \infty$ only the smallest root in absolute value
matters. To select the appropriate root, it suffices to plot the combination
$(t_i - 1/3)/2 \sqrt{-Q}$, $i = 1,2,3$,
versus the variable $x = R/\sqrt{-Q^3}$, which is defined on the interval $[-1,1]$.
This plot is presented in Fig.6 and shows that the smallest root is $t = t_2$.
\begin{figure}[ht]
\begin{center}
\includegraphics*[scale=0.4]{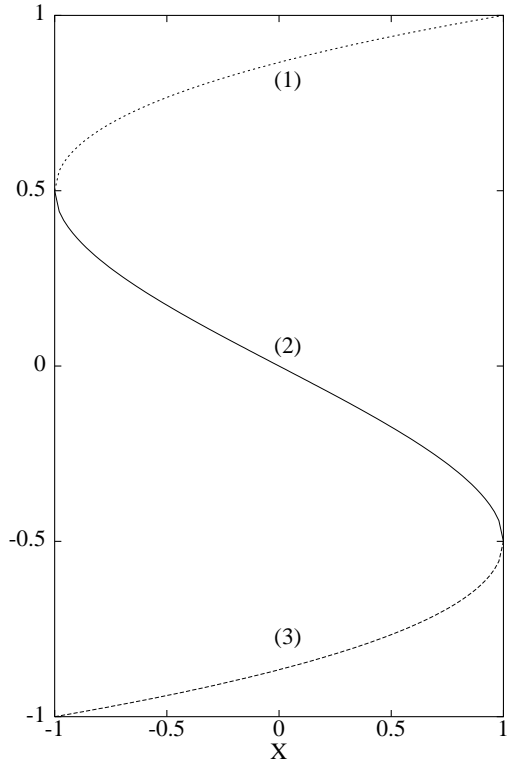}
\end{center}
\caption{\label{Fig6} {\small \sl Plot of $(t_i - 1/3)/2 \sqrt{-Q}$, $i = 1,2,3$,  versus the variable
$X = R/\sqrt{-Q^3}$. The dotted line $(1)$ gives $(t_1 - 1/3)/2 \sqrt{-Q}$, the solid line $(2)$ gives
$(t_2 - 1/3)/2 \sqrt{-Q}$, and the dashed line $(3)$ gives $(t_3 - 1/3)/2 \sqrt{-Q}$, respectively.}}
\end{figure}

Consequently, taking into account that
\begin{equation}
\frac{1 + 3 \sqrt{- Q} \sin\Big(\displaystyle \frac{1}{3}
\arcsin\Big(R/\sqrt{-Q^3}\Big)\Big)}{2 \sin\Big(\displaystyle \frac{\pi}{6} + \frac{2}{3}
\arccos\Big(R/\sqrt{-Q^3}\Big)\Big) - 1} \geq 0,
\end{equation}
is positive definite, we find from eq.(\ref{ann}) that the pressure per site is
given by
\begin{eqnarray}
\label{anni}
\beta P^{(ann)}(p) &=&  \ln\Big(\frac{z (1-p)}{t_2}\Big) = \nonumber\\
&=& - \ln\left(\frac{\Big(1 - 6 \sqrt{-Q} \sin\Big( \displaystyle \frac{1}{3}
\arcsin\Big(R/\sqrt{-Q^3}\Big)\Big)\Big)}{3 z (1-p)}\right),
\end{eqnarray}
which expression determines $P^{(ann)}(p)$ for arbitrary values of the
parameters $p$ and $z$.

Consider now the form of  $P^{(ann)}(p)$ in the limits $p \to 0$
and $p \sim 1$. In the limit of a vanishingly small concentration of
catalytic sites, we have  from eq.(\ref{anni}) that:
\begin{equation}
\beta P^{(ann)}(p) = \ln(1+z) -
\frac{z^2 (2 + z)}{\beta (1 + z)^3} p + {\mathcal O}\Big(p^{2}\Big).
\label{rrrrr}
\end{equation}
Note that the first term in the last equation is, as it should
be,
just the standard Langmuir adsorption isotherm.

In the case $p \sim 1$, we  expand first
$t_2/z (1 -p)$ in powers of $(1 - p)$. This yields,
\begin{equation}
\frac{t_2}{z (1 -p)} = \frac{\sqrt{1 + 4 z} - 1}{2 z} -
\frac{(1 + z) \sqrt{1 + 4 z} - 3 z - 1}{2 z \sqrt{1 + 4 z}} (1 -
p)^2 + {\mathcal O}\Big((1 - p)^3\Big),
\label{fi}
\end{equation}
which expansion, as one can check by comparing the first and the second terms
in eq.(\ref{fi}), makes sense only when $(1 - p) \ll (4/z)^{1/4}$. 
As a matter of fact, $p = 1$ has here a special role, as will be shown in the following.
Then,
we find that the pressure per site is
\begin{equation}
\beta P^{(ann)}(p) =  \ln\Big(\frac{\sqrt{1 + 4 z} + 1}{2 }\Big) +
\frac{4 z^2 + 5 z - 3 z \sqrt{1 + 4 z} + 1 - \sqrt{1 + 4 z} }{(\sqrt{1 + 4 z} - 1) (1 + 4 z)} (1 - p)^2 + {\mathcal O}\Big((1 - p)^3\Big)
\label{j}
\end{equation}

Note now that the first term in eq.(\ref{j}), which defines the disorder-averaged pressure
 per site in the limit $p = 1$,
coincides exactly with the result we obtained earlier in the case of a
regular placement of catalytic sites with the
period $L = 1$ (or $L = 2$), eq.(\ref{i}). As a matter of fact, one
could notice that the partition function in
the annealed-disorder case, eq.(\ref{gen}),
will coincide for $p = 1$ with the partition function for the regular case,
eqs.(\ref{explicit}) and (\ref{L}), by just analysing the behavior of the coefficients $M_m({\it N}_+|N)$,
eqs.(\ref{y}) and (\ref{x}). To show this, we note first that from eqs.(\ref{y}) and (\ref{x}) one has
that $M_m({\it N}_+|N) \equiv 0$ for $m > {\it N}_+$. This is, of course, quite evident, if we
recall that $M_m({\it N}_+|N)$ stands for the number of realizations $\{n_i\}$ having a fixed number
${\it N}_+$ of adsorbed particles and a fixed number $m$ of elementary cells containing one adsorbed particle
surrounded by two vacant sites: hence,  the number of realizations $\{n_i\}$ in which $m$ exceeds
${\it N}_+$ equals zero. Further on, one notices that when $p = 1$, in eq.(\ref{gen})
only the terms $M_{m = {\it N}_+}({\it N}_+|N)$ with ${\it N}_+ \leq {\it N}_- + 1 = N + 1 - {\it N}_+$
matter. Hence  eq.(\ref{gen}) becomes
\begin{equation}
\label{qqq}
\left<Z_N(\zeta)\right> = \sum_{{\it N}_+ =0}^{[(N + 1)/2]}
 M_{{\it N}_+}({\it
N}_+|N) \; z^{{\it N}_+},
\end{equation}
where
\begin{eqnarray}
\label{qqqq}
&&M_{m = {\it N}_+}({\it N}_+|N) = \frac{1}{2 \pi i} {{\it N}_-+1 \choose {\it N}_+}   \oint_{{\cal C}} \frac{d\tau}{\tau}
S_{\tau}^{{\it N}_- + 1 - {\it N}_+}  = \nonumber\\
&=& \frac{1}{2 \pi i} {{\it N}_-+1 -{\it N}_+ \choose {\it N}_+} 
 \oint_{{\cal C}} \frac{d\tau}{\tau} \Big(1 + \sum_{k = 2}^{\infty}
\tau^k \Big)^{{\it N}_- + 1 - {\it N}_+} 
\equiv {{\it N}_-+1 -{\it N}_+ \choose {\it N}_+} 
\end{eqnarray}
On comparing eqs.(\ref{qqq}) and (\ref{qqqq}) with eq.(\ref{pol}), we have in the limit $p = 1$ that the
partition function, (and hence, the pressure per site), in the annealed disorder case coincides with the
partition function of a chain 
on which the
catalytic sites are placed regularly
with period $L = 1$.

Finally, differentiating eq.(\ref{anni}) with respect to the variable $z$ and making use of
eq.(\ref{density}), we find that in the annealed disorder case
the average density of adsorbed particles obeys:
\begin{eqnarray}
\label{density2}
&&n^{(ann)}(p) = 1 + \frac{3 z (1 - p)^2}{1 - 6 \sqrt{- Q}
\sin\Big(\displaystyle \frac{1}{3}
\arcsin\Big(R/\sqrt{-Q^3}\Big)\Big)}  \times \nonumber\\
&\times& \Big[A \sin\Big(\frac{1}{3}
\arcsin\Big(R/\sqrt{-Q^3}\Big)\Big) + \frac{B}{\sqrt{1 + R^2/Q^3}}  \cos\Big(\frac{1}{3}
\arcsin\Big(R/\sqrt{-Q^3}\Big)\Big)\Big],
\end{eqnarray}
where
\begin{equation}
A = \Big( p (3 z p^2 - 2 p (1 + 3 z) + 3 ( 1 + z))\Big)^{-1/2},
\end{equation}
and
\begin{equation}
B = \frac{3}{2} \frac{(2 z p^2 + p (5 - 4 z) + 2 z - 7)}{\Big( 3 z p^2 - 2 p (1 + 3 z)
 + 3 ( 1 + z)\Big)^2}.
\end{equation}
Note that the result in 
eq.(\ref{density2}) differs from the mean-field prediction
$\overline{n} \equiv 0$, which follows from eq.(\ref{no}) for instantaneous reactions, ${\cal K} = \infty$.

In Fig.8 we present a plot 
of $n^{(ann)}(p)$ versus $p$ for different values of the activity $z$. 
In Fig.8 we also compare
the behavior of $n^{(ann)}(p)$ with the behavior found 
in the case of quenched disorder (see the next section). 
In the limits
$p \ll 1$ and $p \sim 1$,  we find that $n^{(ann)}(p)$ follows, respectively,
\begin{equation}
\label{0}
n^{(ann)}(p) = \frac{z}{1 + z} -  \frac{(4 + z) z^2}{(1 + z)^4} p +  {\mathcal O}\left(p^2\right),
\end{equation}
and
\begin{equation}
\label{00}
n^{(ann)}(p) = 1 - \frac{2 z}{1 + 4 z - \sqrt{1 + 4 z}} + \frac{2 z^2}{(1 + 4 z)^{3/2}}  (1 - p)^2 + 
{\mathcal O}\left((1 - p)^3\right).
\end{equation}
Note that the first term in eq.(\ref{0}) is, as it should be, just the Langmuir
adsorbtion isotherm, while the first term in eq.(\ref{00}) coincides with
our earlier result for the average density of adsorbed particles on the periodic chain
with $L = 1$ or $L = 2$, see eq.(\ref{"}). 

Lastly, we analyse the behavior of the compressibility $k_T^{(ann)}(p)$
for the  annealed disorder situation:
\begin{equation}
\beta^{-1} k_T^{(ann)}(p) = \frac{z}{(n^{(ann)}(p))^2} \frac{\partial n^{(ann)}(p)}{\partial z}.
\end{equation} 
We find that $k_T^{(ann)}(p)$   shows the following asymptotical behavior:
For $p \ll 1$
we have
\begin{equation}
\beta^{-1} k_T^{(ann)}(p) = \frac{1}{z} + \frac{z (7 + z)}{(1 + z)^3} p + {\cal O}\Big(p^{2}\Big),
\end{equation}
while for $p \sim 1$ the compressibility obeys
\begin{equation}
\label{comp2}
\beta^{-1} k_T^{(ann)}(p) = \beta^{-1} k_T^{(reg)}(L = 1 \, {\rm or} \, 2) + 
\frac{4 z^2}{(1 + 4 z)^{3/2}} (1 - p)^2 + {\cal O}\Big((1 - p)^{3}\Big).
\end{equation}  
Here $k_T^{(reg)}(L = 1 \, {\rm or} \, 2)$ is as previously defined,  eq.(\ref{comp1}), and represents the compressibility
of a completely catalytic chain. 
Note that the result in eq.(\ref{comp2}) signifies that in the annealed disorder case the compressibility is a 
non-monotonic function of the mean density $p$ 
of  catalytic sites. This can be seen immediately if one notices that, 
first, for any fixed $z$, one has
$k_T^{(reg)}(L = 1 \, {\rm or} \, 2) \geq k_T^{(lan)} = \beta/z$, (or in other words, that for any fixed $z$
the compressibility of a non-catalytic (Langmuir) system is always 
smaller than the compressibility of a completely catalytic system)
and second, that $k_T^{(ann)}(p)$ approaches $k_T^{(reg)}(L = 1 \, {\rm or} \, 2)$ from above, since the
 function $ 4 z^2/(1 + 4 z)^{3/2}$ is always positive.

We close this section with some comments concerning the 
large-$z$ behavior of $P^{(ann)}(p)$ and $n^{(ann)}(p)$. 
As a matter of fact, it appears that
in the $annealed $ disorder case the large-$z$ behavior of $P^{(ann)}(p)$ and $n^{(ann)}(p)$
for $p$ arbitrarily close but strictly less than unity is completely different from the large-$z$ behavior of these parameters
in the case when $p \equiv 1$. This implies, of course, 
that $p = 1$ is a special point. More specifically, we find that for
 $z \gg (1 - p)^{-2}$ the disorder-averaged pressure per site
obeys
\begin{equation}
\beta P^{(ann)}(p) = \ln(z) + \ln(1 - p) + \frac{1}{(1 - p) z} - 
\frac{(1 - 3 p)}{(1 - p)^3 z^2} + {\cal O}\Big(\frac{1}{z^3}\Big),
\end{equation} 
which implies that the mean density follows:
\begin{equation}
n^{(ann)}(p) = 1 - \frac{1}{(1 - p) z} + \frac{(1 - 3 p)}{(1 - p)^3 z^2} + {\cal O}\Big(\frac{1}{z^3}\Big).
\end{equation} 
This asymptotic behavior should be contrasted to the asymptotic behavior
which holds in the $p \equiv 1$ case. When $z \to \infty$, we find from eqs.(\ref{i}) and (\ref{"}) the following results:
\begin{equation}
\label{4i}
\beta P^{(reg)}(L = 1 \, {\rm or} \, 2) = \frac{1}{2} \ln(z) + \frac{1}{2 \; z^{1/2}} - \frac{1}{48 \;
z^{3/2}} + \frac{3}{1280 \; z^{5/2}} + {\cal O}\Big(\frac{1}{z^{7/2}}\Big),
\end{equation}
and
\begin{eqnarray}
n^{(reg)}(L = 1 \, {\rm or} \, 2) = \frac{1}{2} - \frac{1}{4 \; z^{1/2}} + \frac{1}{32 \; z^{3/2}} -
\frac{3}{512 \; z^{5/2}} + {\cal O}\left(\frac{1}{z^{7/2}}\right).  
\end{eqnarray} 
This signifies, in particular, 
that for $p$ arbitrarily close but not equal to unity, the mean density is equal to
$1$ as $z = \infty$, while for $p$ strictly equal to unity the mean density $n^{(reg)}(L = 1 \, {\rm or} \, 2) = 1/2$.
On physical grounds, such a behavior can be understood as follows. In the annealed disorder case, 
instead of averaging the logarithm of the partition function in eq.(\ref{partition}),
we average
the
partition function itself and thus operate with an effective, "annealed" partition function in eq.(\ref{zu}).
Here, the strict constraint that no two particles can occupy 
simultaneously neighboring sites if at least one of them sits on the catalytic site, is replaced
by a more tolerant condition (see, eq.(\ref{psipsi})), which allows for such pairs to be present
at any site, while a penalty 
of $2 \ln(1 - p)$ is to be paid. For any finite $p < 1$ such a penalty can be overpassed by increasing 
the chemical potential. Thus for $\beta \mu \gg - 2 \ln(1 - p)$ (or, equivalently, for $z \gg (1 - p)^{-2}$)
one expects essentially the same behavior regardless of the value of $p$, and, in particular,
that $n^{(ann)}(p) \to 1$ as $z \to \infty$. On the other hand, for $p \equiv 1$ the penalty for
having a pair of particles occupying neighboring sites becomes infinitely large and can not 
be compensated by any 
increase of the chemical potential.     
The behavior of $n^{(ann)}(p)$ as a function of $z$ for different values 
of $p$ is depicted in Fig.9 and compared with the behavior obtained for it in
the case of quenched disorder.

\section{Random placement of catalytic sites. Quenched disorder.}

We finally turn to the most challenging situation - 
the case of quenched randomness in the placement of catalytically active sites. We
begin by introducing one auxiliary function. Consider a chain of 
length $N$ which contains a fixed number $N - N_{nc}$ of
catalytic and hence, a fixed number $N_{nc}$ of non-catalytic sites, 
the latter being placed at the positions $\{X_n\}$,
$n = 1, 2, \ldots , N_{nc}$. We denote the partition function of such a chain 
as $Z_N(\{X_n\})$. Evidently,
$Z_N(\{X_n\})$ obeys eq.(\ref{partition}) and eq.(\ref{partition2}) with
\begin{equation}
\zeta_i = \left\{\begin{array}{ll}
0,     \mbox{if the $i \in \{X_n\}$, } \nonumber\\
1,     \mbox{ otherwise.}
\end{array}
\right.
\end{equation}
Then, the logarithm of the partition function in eq.(\ref{partition}), averaged over all realizations of the quenched
random variable $\{\zeta_i\}$, can be formally written as
\begin{equation}
\label{ququ}
\left<  \ln Z_N(\zeta)\right> \; = \; \sum_{N_{nc} = 0}^N p^{N - N_{nc}} (1 - p)^{N_{nc}} \sum_{\{X_n\}} \ln Z_N(\{X_n\}),
\end{equation}
where the sum with the subscript $\{X_n\}$ signifies that the summation extends over all possible placements of $N_{nc}$
non-catalytic sites on an $N$-chain.

Next, similarly to the approach used in the previous section, we introduce a set of $N_{nc} + 1$ intervals
$\{l_{n}\}$ determined by consecutive non-catalytic sites, such that $l_n = X_n - X_{n -1}$ (with $X_0 = 0$) and
$l_{N_{nc} + 1} = N + 1 - X_{N_{nc}}$.  That is, the first interval extends 
from the boundary (non-catalytic, unoccupied)
site $i = 0$ to its closest non-catalytic neighboring 
site, the second interval extends from this non-catalytic site to the
following one, and so forth, while the interval $l_{N_{nc} + 1}$ goes from 
the last non-catalytic site inside the chain to the
boundary site $i = N + 1$. In terms of these intervals,  eq.(\ref{ququ}) can be rewritten as
\begin{equation}
\label{ququq}
\left<  \ln Z_N(\zeta)\right> \; = \; \sum_{N_{nc} = 0}^N p^{N - N_{nc}} (1 - p)^{N_{nc}} \sum_{\{l_{n}\}}
\ln Z_N(\{l_{n}\}),
\end{equation}
where $Z_N(\{l_{n}\})$ stands now for the partition function $ Z_N(\{X_n\})$ in which we have just
expressed the
positions of the non-catalytic sites using the set $\{l_{n}\}$, while the summation with the sign
$\{l_{n}\}$ denotes now the summation over all possible solutions of the equation, analogous to
eq.(\ref{dio}),
\begin{equation}
\label{dina}
l_1 + l_2 + l_3 + \ldots + l_{N_{nc} + 1} = N + 1,
\end{equation}
where each $l_i \geq 1$.

For each given set $\{l_{n}\}$ of intervals we have that the partition function of an
$N$-chain decomposes into that of smaller clusters,
\begin{equation}
\label{deco}
Z_N(\{l_{n}\}) = Z_1^{{\cal N}_1(\{l_{n}\}|N)} \;
Z_2^{{\cal N}_2(\{l_{n}\}|N)} \; Z_3^{{\cal N}_3(\{l_{n}\}|N)} \; \ldots \;
Z_N^{{\cal N}_N(\{l_{n}\}|N)},
\end{equation}
where $Z_K$, $(K = 1, 2, \ldots ,N)$, is the partition function of the $K$-cluster, which obeys
eqs.(\ref{explicit}) and (\ref{L}) or (\ref{pol}) (with $N$ replaced by $K$), while
${\cal N}_K(\{l_{n}\}|N)$ denotes the $\{l_{n}\}$-realization  dependent number of
$K$-clusters in an $N$-chain with $N_{nc}$ non-catalytic sites. Evidently,
for each realization  $\{l_{n}\}$ these numbers ${\cal N}_K(\{l_{n}\}|N)$ obey
\begin{equation}
\label{cons}
{\cal N}_1(\{l_{n}\}|N) + 2 \; {\cal N}_2(\{l_{n}\}|N) + 3 \; {\cal N}_3(\{l_{n}\}|N) + 
\; \ldots \; + N \; {\cal N}_N(\{l_{n}\}|N) = N
\end{equation}

Note now that we have previously defined a 
$K$-cluster as being 
the set of all sites connected
consecutively by arcs (see Fig.1). An equivalent 
definition of the $K$-cluster, 
which uses now the language of the intervals
connecting consecutive non-catalytic sites 
is as follows:
an interconnected  $K$-cluster, whose partition function is determined by
eqs.(\ref{explicit}) and (\ref{L}) or (\ref{pol}) (with $N$ replaced by $K$),
is a subset of $n$ ($n \leq [(K-1)/2]$) consequitive intervals $l_{r + 1}, l_{r + 2},
l_{r + 3}, \ldots , l_{r + n}$ from the entire set $\{l_{n}\}$, 
where all intervals a) are
greater than unity, b) 
obey the conservation law $l_{r + 1} +
l_{r + 2} + \ldots + l_{r + n } = K - 1$, and c) are
 necessarily  bounded by two intervals ($l_r$ and $l_{r + n + 1}$) of length unity.
\begin{figure}[ht]
\begin{center}
\includegraphics*[scale=1.0]{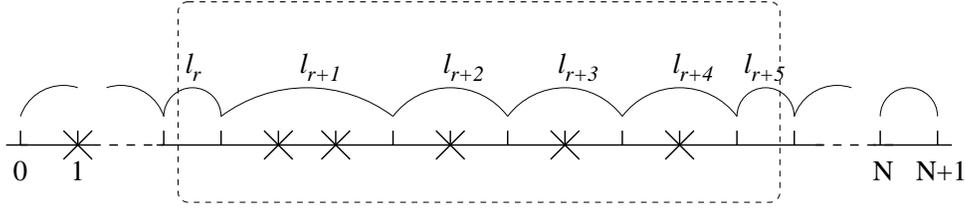}
\end{center}
\caption{\label{Fig7} {\small \sl Example of a 10-cluster containing $4$ inner intervals and $5$ non-catalytic sites.}}
\end{figure}
As we have already remarked, 
the latter condition 
implies that two pairs of non-catalytic sites appear at the extremities of the 
segment containing the $K$ sites
which automatically decomposes the chain into three independent parts. On the other hand, the 
condition that
all intervals $l_k$, $k = r + 1, r + 2, \ldots , r + n$, are greater than unity insures that the
$K$-cluster is interconnected and does not break up into smaller subunits (see Fig.7). Consequently, we have
from eq.(\ref{deco}) that
\begin{equation}
\label{decom}
\ln Z_N(\{l_{n}\}) = \sum_{K = 1}^N {\cal N}_K(\{l_{n}\}|N) \; \ln Z_K,
\end{equation}
and eq.(\ref{ququq}) becomes
\begin{equation}
\label{quququq}
\left<  \ln Z_N(\zeta)\right> \; = \; \sum_{N_{nc} = 0}^N p^{N - N_{nc}} (1 - p)^{N_{nc}}
 \sum_{K = 1}^N N_K(N_{nc}|N) \; \ln Z_K,
\end{equation}
where now $N_K(N_{nc}|N)$ reads
\begin{equation}
N_K(N_{nc}|N) = \sum_{\{l_{n}\}} {\cal N}_K(\{l_{n}\}|N),
\end{equation}
and hence, defines the total number of $K$-clusters
in all realizations of the $N$-chain with a fixed number
$N_{nc}$ of non-catalytic sites.

The disorder-averaged pressure
 per site in the $N$-chain with random, quenched placements of catalytic sites is then
given by
\begin{equation}
\label{pree}
\beta P^{(quen)}(p)  = \lim_{N \to \infty} \frac{1}{N} \sum_{K = 1}^N w_{K,N}(p) \ln Z_K,
\end{equation}
where now $w_{K,N}(p)$ is the statistical weight
of the $K$-clusters in the $N$-chain; $w_{K,N}(p)$ obeys
\begin{equation}
w_{K,N}(p) =  \sum_{N_{nc} = 0}^N p^{N - N_{nc}} \; (1 - p)^{N_{nc}} \; N_K(N_{nc}|N).
\end{equation}
Below we determine $w_{K,N}(p)$ explicitly.

\subsection{Calculation of the weights $w_{K,N}(p)$.}

To fix the ideas, we start with the trivial case of $(K = 1)$- and $(K = 2)$-clusters.
Consider a given realization $\{l_{n}\}$ of intervals. As may readily notice, a
$(K = 1)$-cluster, or "a free site" in the terminology of the section 3, appears
as soon as one has three consecutively placed non-catalytic sites. In other words,
such a site appears
as soon as any two consecutive intervals $l_r$ and $l_{r+ 1}$
are both equal to unity. Consequently, the number
${\cal N}_1(\{l_{n}\}|N)$ of $(K = 1)$-clusters in a given realization of
the $N$-chain with
$N_{nc}$ non-catalytic sites can be written as follows:
\begin{equation}
{\cal N}_1(\{l_{n}\}|N) = \sum_{r = 1}^{N_{nc} } \delta(l_r,1) \; \delta(l_{r + 1},1)
\end{equation}
Then, using the definition of the lattice delta-function
in eq.(\ref{delta}), we have that the total number  $ N_1(N_{nc}|N)$ of
$(K=1)$-clusters
in all realizations of the $N$-chain with a fixed number
$N_{nc}$ of non-catalytic sites obeys:
\begin{eqnarray}
N_1(N_{nc}|N) &=&   \sum_{r = 1}^{N_{nc} } \sum_{\{l_{n}\}}
\delta(l_r,1) \; \delta(l_{r + 1},1) = 
 \frac{N_{nc}}{2 \pi i} \sum_{\{l_{n}\}} \; \oint_{{\cal C}} \frac{d\tau}{\tau} \;
\tau^{ (\sum_{r = 1}^{N_{nc} - 1} l_r - N + 1) } = \nonumber\\
&=&  \frac{N_{nc}}{2 \pi i}  \; \oint_{{\cal C}}  \frac{d\tau}{\tau}  \;
\Big(\frac{\tau}{1 - \tau}\Big)^{N_{nc} - 1} \; \tau^{-
(N - 1)},
\end{eqnarray}
which yields, using the expansion
\begin{equation}
\label{expansion}
\frac{1}{(1 - \tau)^{N_{nc} - 1}} =
\sum_{n = N_{nc} - 2}^{\infty} {n \choose N_{nc}-2} \; \tau^{\displaystyle (n - N_{nc} + 2)},
\end{equation}
the following result:
\begin{equation}
N_1(N_{nc}|N) =  N_{nc}  \; {N - 2 \choose N_{nc}-2}  \; \times \; \left\{\begin{array}{ll}
1,     \mbox{if the $N_{nc} \geq 2$, } \nonumber\\
0,     \mbox{ otherwise.}
\end{array}
\right.
\end{equation}
Consequently, the weight $w_{1,N}(p)$ of the $(K = 1)$-clusters is given by
\begin{eqnarray}
\label{107}
w_{1,N}(p) &=& \sum_{N_{nc} = 2}^N p^{N - N_{nc}} \; (1 - p)^{N_{nc}}
\; N_{nc} \; {N - 2 \choose N_{nc}-2}  = \nonumber\\
&=& N \; (1 - p)^3 \; + \; 2 \; (1 - p)^2 \; p.
\end{eqnarray}
Next, we turn to calculation of $w_{2,N}$ describing the weight of
the $(K = 2)$-clusters. Two such clusters, as
we have already remarked, may only appear on the chain boundaries in the case when the
sites $i = 1$ or $i = N$ (or both)
are catalytic, while two pairs of neighboring sites $i = 2, 3$ and
$i = N-1, N-2$ are non-catalytic. Consequently, the number of $(K = 2)$-clusters in a
 given realization of an $N$-chain
with $N_{nc}$ non-catalytic sites is given by
\begin{equation}
{\cal N}_2(\{l_{n}\}|N) =  \delta(l_1,2) \; \delta(l_{2},1) + \delta(l_{N_{nc} + 1},2) \;
\delta(l_{N_{nc}},1).
\end{equation}
Hence,
\begin{equation}
N_2(N_{nc}|N) = \frac{1}{\pi i}  \sum_{\{l_{n}\}} \; \oint_{{\cal C}} \frac{d\tau}{\tau} \;
\tau^{ (\sum_{r = 1}^{N_{nc} - 1} l_r - N + 2)} =  \frac{1}{\pi i}   \; \oint_{{\cal C}}  \frac{d\tau}{\tau} \;
\Big(\frac{\tau}{1 - \tau}\Big)^{N_{nc} - 1} \; \tau^{-
(N - 2)},
\end{equation}
and, making use of the expansion in eq.(\ref{expansion}), we obtain
\begin{equation}
N_2(N_{nc}|N) = 2 \;  {N - 3 \choose N_{nc}-2} \; \times \; \left\{\begin{array}{ll}
1,     \mbox{if the $2 \leq N_{nc} \leq N - 1$, } \nonumber\\
0,     \mbox{ otherwise.}
\end{array}
\right.
\end{equation}
Finally, we get
\begin{equation}
w_{2,N} =  2 \; (1 - p)^2 \; p.
\end{equation}
Now, in contrast to the very simple $(K = 1)$- and $(K=2)$-clusters, clusters of
larger size may be composed of several types of intervals. Let ${\cal N}_K^{(n)}(\{l_{n}\}|N)$ 
denote the number of $K$-clusters composed of $n$ intervals
in a given realization of an $N$-chain containing 
 exactly $N_{nc}$ non-catalytic sites. This number can be written down explicitly as
\begin{equation}
{\cal N}_K^{(n)}(\{l_{n}\}|N) = J^{(S)}_{(n)}(\{l_{n}\}|K|N) + J^{(B)}_{(n)}(\{l_{n}\}|K|N), 
\end{equation}
where
\begin{equation}
J^{(S)}_{(n)}(\{l_{n}\}|K|N) = 2\; \Big(\prod_{i =
1}^{n} \delta(l_i \geq 2) \Big) \; \delta(l_{n+ 1},1) \;
\delta(l_1 + l_2 + \;
\ldots \; + l_n, K)
\end{equation}
denotes the contribution from the $K$-clusters starting from either boundary site, i.e. 
"surface" $K$-clusters, 
while
\begin{eqnarray}
J^{(B)}_{(n)}(\{l_{n}\}|K|N) &=&  \sum_{r = 1}^{N_{nc} - n} \delta(l_r, 1) \; \Big(\prod_{i = r
+ 1}^{n + r} \delta(l_i \geq 2) \Big) \times \nonumber\\
&\times&  \delta(l_{r + n + 1},1)
\; \delta(l_{r+1} + l_{r + 2} + \; \ldots \; + l_{r + n}+1,
K)
\end{eqnarray}
represents the contribution of the 
"bulk" $K$-clusters, i.e. such $K$-clusters which are entirely inside the chain and
do not include any of the boundary sites.

Summing 
${\cal N}_K^{(n)}(\{l_{n}\}|N)$
over all the interval realizations $\{l_{n}\}$
obeying the conservation law in eq.(\ref{dina}), 
and next, performing in the result summation over all
possible numbers $n$ of 
subintervals 
in a $K$-cluster,
we find that
for $K\neq 1$ and $K\neq N$, the total
 weight of $K$-clusters is given by
\begin{eqnarray}
\label{w}
w_{K,N}(p)&=&p^{(K-1)/2} \; (1-p)^{(K+3)/2} \Big\{2F_{K}\left(\sqrt{\frac{p}{1-p}}\right) + \nonumber\\
&+&(1-p)(N-K-1)F_{K-2}\left(\sqrt{\frac{p}{1-p}}\right)\Big\},
\end{eqnarray}
while for $K = N$ one has
\begin{eqnarray}
\label{ww}
&&w_{N,N}(p)=p^{N/2}\;(1-p)^{N/2}\;\Big\{
2F_{N-1}\left(\sqrt{\frac{p}{1-p}}\right)+\nonumber\\
&+&\sqrt{\frac{p}{1-p}} F_{N-2}\left(\sqrt{\frac{p}{1-p}}\right)+
\sqrt{\frac{p}{1-p}} F_{N}\left(\sqrt{\frac{p}{1-p}}\right)\Big\}.
\end{eqnarray}
Details of these calculations are presented in Appendix B. 

\subsection{The thermodynamic limit $N \to \infty$.}

Now, having calculated the weights 
$w_{K,N}(p)$ of $K$-clusters explicitly,
we may rewrite
eq.(\ref{pree}) 
as the sum of three contributions
\begin{equation}
\beta P^{(quen)}(p) = \beta  P^{(quen)}_1(p) + \beta  P^{(quen)}_2(p) + \beta  P^{(quen)}_3(p),
\end{equation}
where
the first term $\beta  P^{(quen)}_1(p)$ accounts for the contribution of $(K =1)$-clusters,
\begin{equation}
\beta  P^{(quen)}_1(p) = \lim_{N\to+\infty}\left(\frac{1}{N}
w_{1,N}\ln{Z_1}\right),
\end{equation}
the second term denotes the contribution of a single spanning $N$-cluster,
\begin{equation}
\beta  P^{(quen)}_2(p) = \lim_{N\to+\infty}\left(\frac{1}{N}
w_{N,N}\ln{Z_N}\right),
\end{equation}
while $\beta  P^{(quen)}_3(p)$ takes into account all remaining possible $K$-clusters,
\begin{equation}
\label{third}
\beta  P^{(quen)}_3(p) = \lim_{N\to+\infty}\left(\frac{1}{N}\sum_{K=2}^{N-1}\;w_{K,N}(p)\;\ln{Z_K} \right).
\end{equation}
In all these equations $Z_K$ stands for the partition function of the corresponding $K$-cluster,
which has been defined previously in eqs.(\ref{explicit}) and (\ref{L}).

Now, using the results in eq.(\ref{107}), we readily find that $\beta  P^{(quen)}_1(p)$ obeys
\begin{equation}
\label{single}
\beta  P^{(quen)}_1(p) \equiv (1-p)^3 \ln(1+z).
\end{equation}
Turning next to the contribution due to a single
spanning $N$-cluster, we have that it is given explicitly by
\begin{eqnarray}
\label{tt}
\beta  P^{(quen)}_2(p) &=& \lim_{N \to \infty} \Big(\frac{\displaystyle p^{N/2} (1 - p)^{N/2}}{N} \Big\{2
F_{N-1} \left(\sqrt{\frac{p}{1 - p}}\right) + \nonumber\\
&+& \sqrt{\frac{p}{1 - p}} F_{N-2} \left(\sqrt{\frac{p}{1 - p}}\right) + 
\sqrt{\frac{p}{1 - p}} F_{N} \left(\sqrt{\frac{p}{1 - p}}\right)\Big\} \times \nonumber\\
&\times& \Big[\ln{\cal L}_N + \ln\Big(\frac{z^2 \tau_1^2}{\sqrt{1 + 4 z}}\Big) - N \ln(\tau_2)\Big]\Big).
\end{eqnarray}
Taking into account the explicit representation of the Fibonacci polynomials in 
eq.(\ref{fibo}), we notice that for $p \geq 1/2$ their growth is suppressed 
by the exponentially vanishing factor $(1 - p)^{N/2}$ in the first line of eq.(\ref{tt}). 
On the other hand, for $p < 1/2$, the vanishing factor $p^{N/2}$ in the first line of eq.(\ref{tt})
controls the large-$N$ behavior. Consequently, as could be expected on physical grounds,
the contribution of a single spanning $N$-cluster is 
exactly equal to zero in the thermodynamic limit.    

Consider now the form of  $\beta P^{(quen)}_3(p)$ in
 eq.(\ref{third}). Taking into account the explicit
form of $Z_N$ in eqs.(\ref{explicit}) and (\ref{L}), 
and expanding ${\cal L}_N$ in eq.(\ref{L}) in Taylor
series in powers of $(\tau_2/\tau_1) < 1$,  
we may rewrite $\beta P^{(quen)}_3(p)$ in eq.(\ref{third}) as 
\begin{eqnarray}
\label{calcul}
\beta P^{(quen)}_3(p) &=& \lim_{N \to \infty} \left( 
\frac{1}{N}\left(\sum_{K=2}^{N-1}w_{K,N}(p)\right)\ln\Big(\frac{z^2 \tau_1^2}{\sqrt{1 + 4 z}}\Big)\;-\;
\frac{1}{N}\left(\sum_{K=2}^{N-1}K \; w_{K,N}(p)\right)\ln{\tau_2} - \right. \nonumber\\
&-&\;\left. \frac{1}{N}\sum_{K=2}^{N-1}w_{K,N}(p)\sum_{n=1}^\infty\frac{(-1)^{n
K}}{n}\left(\frac{\tau_2}{\tau_1}\right)^{n (K + 2)} \right).
\end{eqnarray}
Evaluation of the sums entering  
eq.(\ref{calcul}) is rather cumbersome
and we present the details of such calculations 
in Appendix C. 

Taking into account both the contribution of the $(K = 1)$-clusters in eq.(\ref{single})
and that of $\beta P^{(quen)}_3(p)$ in eq.(\ref{calcul}) (see Appendix C)
we arrive 
at the desired explicit expression describing the disorder-averaged pressure
$P^{(quen)}(p)$ per site
in the quenched disorder case:
\begin{eqnarray}\label{resultat}
\beta P^{(quen)}(p) &=& (1-p)^3\ln(1+z) + p \Big(5 - 7 p + 3 p^2\Big) \ln\left(\frac{\sqrt{1 + 4 z} + 1}{2}\right)
- \frac{p (1 - p)^2}{2} \ln(1 + 4 z) + \nonumber\\
 &+& p (1-p)^4 \; \sum_{n=1}^\infty \frac{(-1)^n}{n} 
\displaystyle \frac{\left(\tau_2/\tau_1\right)^{5 n}}{1-p (-1)^n \left(\tau_2/\tau_1\right)^n-p(1-p)
 \left(\tau_2/\tau_1\right)^{2n}},
\end{eqnarray}
which can be reformulated, (by expanding the denominator 
in elementary fractions), in the following form: 
\begin{eqnarray}\label{resultat integral}
\beta P^{(quen)}(p) &=& (1-p)^3\ln(1+z) + p \Big(5 - 7 p + 3 p^2\Big) \ln\left(\frac{\sqrt{1 + 4 z} + 1}{2}\right)
- \frac{p (1 - p)^2}{2} \ln(1 + 4 z) - \nonumber\\
 &-& \frac{p(1-p)^4}{\sqrt{p(4-3p)}}
\sum_{m=0}^\infty\left(\frac{1}{X_+^m}-\frac{1}{X_-^m}\right)\ln\left(1-(-1)^{m+1}\Big(\frac{\tau_2}{\tau_1}\Big)^{m+5}\right),
\end{eqnarray}
where the $X_{\pm}$ are given by
\begin{equation}\label{def de X+-}
X_{\pm}=\frac{1}{2p(1-p)}\left[-p\pm\sqrt{p(4-3p)}\right].
\end{equation}
We note here that the first correction to the thermodynamic limit result in eqs.(\ref{resultat}) 
or (\ref{resultat integral}), (which shows how fast the thermodynamic limit is approached  with respect to the chain's length),
should be proportional to the first inverse power of $N$, 
as follows from the expansion eq.(\ref{inf}) (see Appendix C).

\begin{figure}[ht]
\begin{center}
\includegraphics*[scale=0.4,angle=-90]{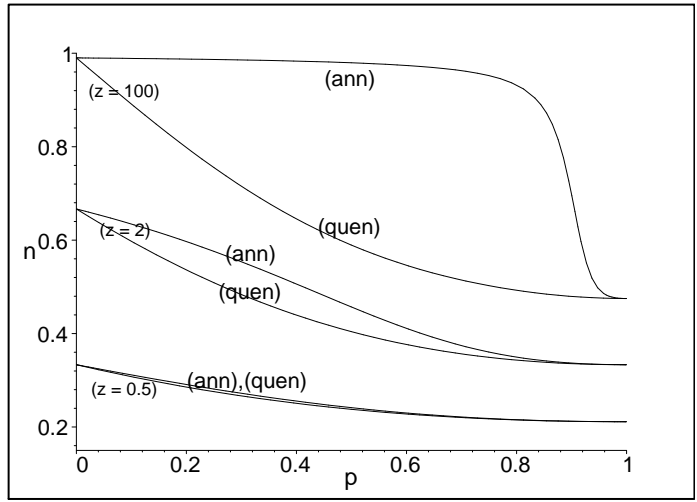}
\caption{\label{Fig8} {\small \sl Mean particle density $n$
 versus mean density  $p$ of catalytic sites for different values of $z = \exp(\beta \mu)$. 
Curves with signs
(ann) and (quen)  
depict the behavior of the mean density for annealed and 
quenched random distributions, respectively.  } }
\end{center}
\end{figure}

We now consider the asymptotical behavior of $P^{(quen)}(p)$ for different limiting cases. 
In the asymptotic limit $p \to 0$ we find from
eq.(\ref{resultat}) that
\begin{equation}
\beta P^{(quen)}(p) = \ln\left(1 + z  \right) +  
\ln\left(\frac {1+3\,z+z^{2}}{ \left( 1+z \right) ^{3}} \right) p + {\cal O}(p^2),
\end{equation}
where the first term represents the Langmuir pressure, and the second one - the small-$p$ correction to it. 
Note that already the first correction term differs significantly from the first correction term to $P^{(ann)}(p)$
found in the annealed disorder case, eq.(\ref{rrrrr}). Next, in the limit $p \sim 1$, we find
that  $P^{(quen)}$ obeys
\begin{equation}
\beta P^{(quen)}(p) = \ln \left(\frac {\sqrt {1+4 z}+1}{2} \right) +
\ln\left( \frac{\Big(1+\sqrt {1+4\,z}\Big)^2}{4\sqrt {1+4\,z}} \right) \left(1 - p \right)^{2}+ 
 {\cal O}\left((1 - p)^3\right),
\end{equation}
in which expansion the second term is also different from 
the one obtained in the annealed disorder case, eq.(\ref{j}).

\begin{figure}[ht]
\begin{center}
\includegraphics*[scale=0.4,angle=-90]{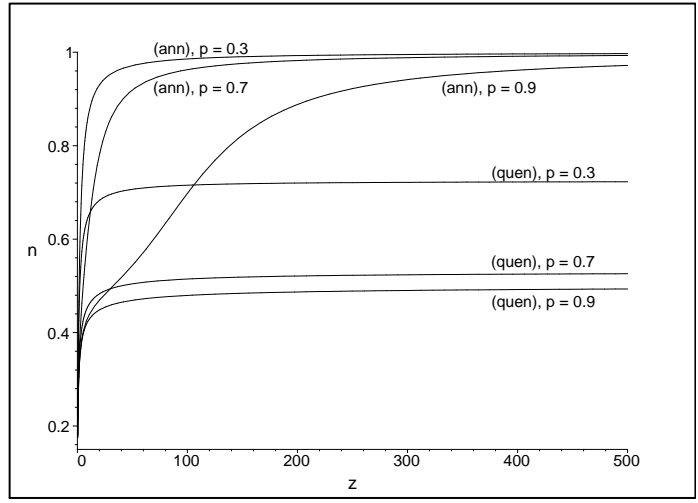}
\caption{\label{Fig9} {\small \sl
Mean particle density $n$ in the annealed and quenched disorder cases, eqs.(\ref{density2}) 
and (\ref{density3}),
 versus the activity $z$ for different values of $p$. 
The notations are the same as in Fig.8.} }
\end{center}
\end{figure}

Now, turning to the analysis of the large-$z$ behavior of $P^{(quen)}$ we notice that 
 the behavior differs completely for $p \equiv 1$ and for $p < 1$, which signifies that
here, (as in the annealed disorder case), $p = 1$ is a special point. 
In the case $p \equiv 1$ we have that on the righthand side of eq.(\ref{resultat})
all terms except for the second one vanish and hence,  $P^{(quen)}(p) = P^{(reg)}(L = 1 \, {\rm or} \, 2)$, eq.(\ref{i}).
Consequently, for $p \equiv 1$ the large-$z$ behavior of $P^{(quen)}(p)$ obeys eq.(\ref{4i}), 
similar to $P^{(reg)}(L = 1 \, {\rm or} \, 2)$ and
to $P^{(ann)}(p)$, which is, of course, not surprising. On the other hand, for $p < 1$ the large-$z$ behavior is 
rather different from the behavior observed in the annealed disorder case. Here, we find for $z \gg (1 - p)^{-2}$
that $P^{(quen)}(p)$ obeys
\begin{eqnarray}
\beta P^{(quen)}(p) =  \frac {\left( p^{2}-p+1 \right) }{p^{2}} \ln\left( z \right) + {\cal O}\left(1\right),
\end{eqnarray}
i.e. in the quenched disorder case 
the prefactor in the leading $z$-term depends on $p$, while in the annealed disorder case this prefactor
was found to be independent of $p$, which caused a rather strange behavior of the mean particle density.

Differentiating eq.(\ref{resultat integral}), we find that the particle density in the quenched disorder case
is explicitly given by
\begin{eqnarray}
\label{density3}
&&n^{(quen)}(p)= (1-p)^3\frac{z}{1+z} + p \Big(5 - 7 p + 3 p^2\Big) \frac{2 z}{1 + 4 z + \sqrt{1 + 4z}}
- \frac{2 p (1 - p)^2 z}{1 + 4 z}  - \nonumber\\
 &-&\frac{4 p(1-p)^4 z}{\sqrt{p(4-3p) (1 + 4 z)} \Big(1 + \sqrt{1 + 4 z}\Big)^2}
\sum_{m=0}^\infty\left(\frac{1}{X_+^m}-\frac{1}{X_-^m}\right)
\frac{\displaystyle (m + 5) \Big(-\frac{\tau_2}{\tau_1}\Big)^{m+4}}{\displaystyle 
\left(1 - \Big(-\frac{\tau_2}{\tau_1}\Big)^{m+5}\right)}.
\end{eqnarray}
We note that again 
this result differs 
considerably from the mean-field prediction $\overline{n} \equiv 0$  
(eq.(\ref{no}) 
with ${\cal K} = \infty$).

From  eq.(\ref{density3}) we find then that 
the asymptotic behavior of $n^{(quen)}(p)$ in the small-$p$ limit
obeys
\begin{eqnarray}
n^{(quen)}(p)=\frac{z}{1+z} - \frac {z^{2} \left( 4+z \right) }{ \left( 1+z
 \right)  \left( 1+3\,z+z^{2} \right)} p + {\cal O}(p^2),
\end{eqnarray}
while in the limit $p \sim 1$ it follows
\begin{equation}
n^{(quen)}(p)=n^{(reg)}(L=1 \, {\rm or} \, 2) + {\frac {4 {z
}^{2}}{ \left( 1+4\,z \right)  \left( 1+2\,z+
\sqrt {1+4\,z} \right) }} 
(1 - p)^2+ {\cal O}\left((1 - p)^3\right),
\end{equation}
in which equations the first corrections to the Langmuir and 
the regular cases 
depend very 
differently on the activity 
$z$ when compared to the annealed disorder case. 
Note also that, as depicted in Fig.8, for any fixed $z$ the mean particle densities 
in the annealed and in the quenched disorder
cases show a completely different 
behavior as functions of the mean density $p$. 
The difference becomes more pronounced
with increasing $z$.

Note that the catalytic efficiency in the annealed disorder case 
turns out to be lower than in the quenched case, as can be inferred from the fact that the
 $A$ particle density is always higher in the former case (see Fig.9).

\begin{figure}[!ht]
\begin{center}
\begin{minipage}[c]{.5\textwidth}
\includegraphics[width=.7\textwidth,angle=-90]{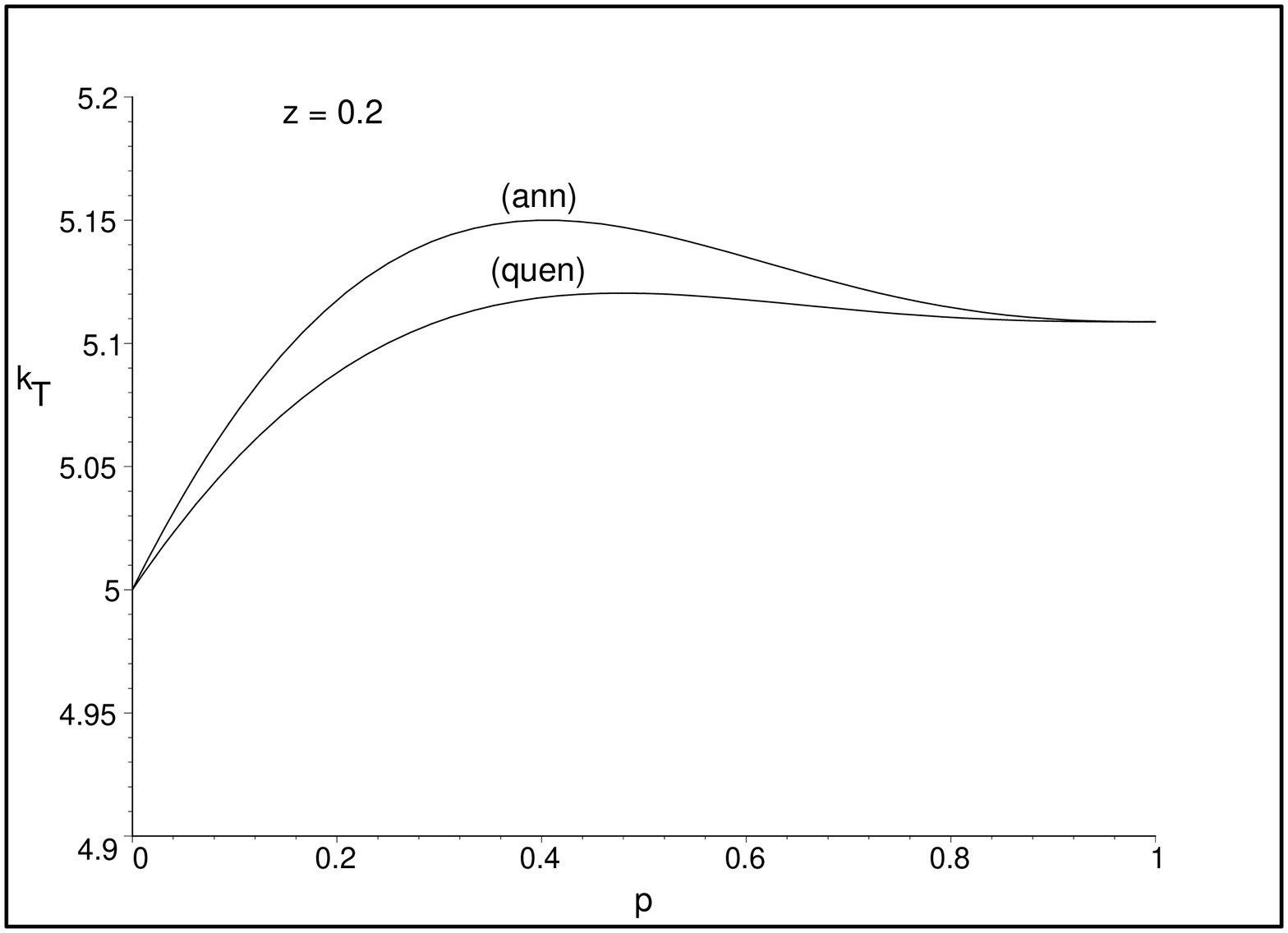}
\end{minipage}%
\begin{minipage}[c]{.5\textwidth}
\includegraphics[width=.7\textwidth,angle=-90]{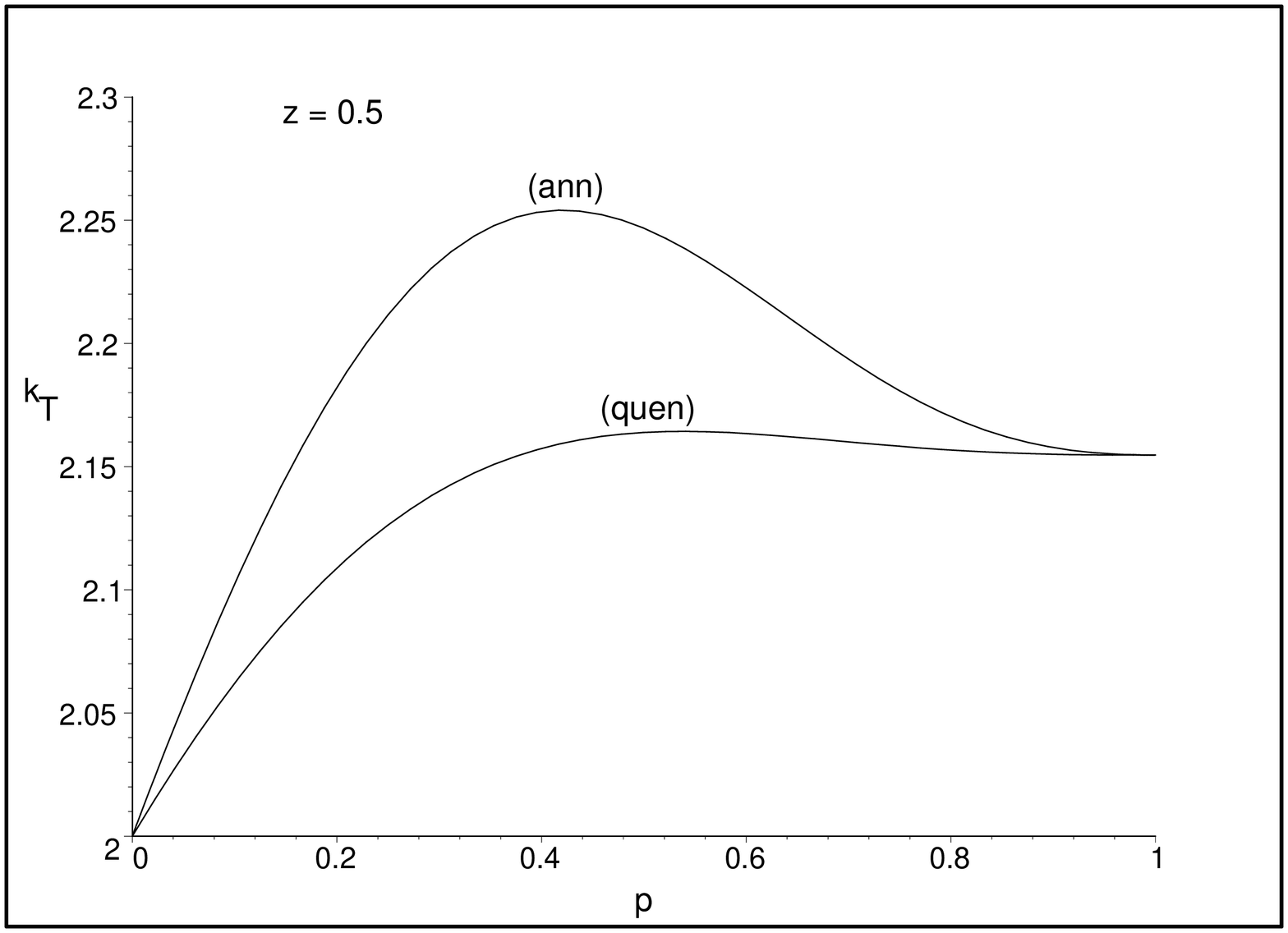}
\end{minipage}
\end{center}
\begin{center}
\begin{minipage}[c]{.5\textwidth}
\includegraphics[width=.7\textwidth,angle=-90]{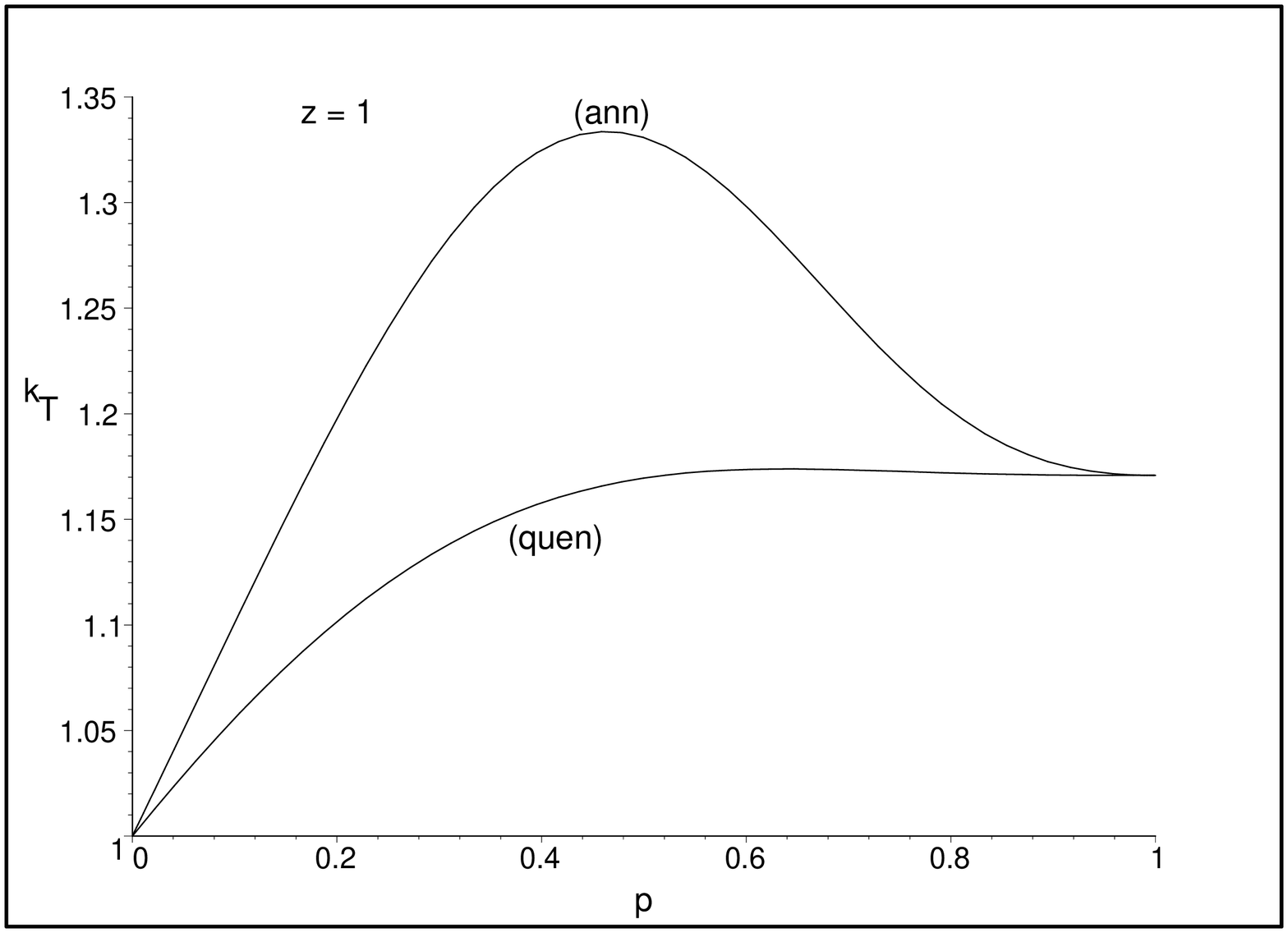}
\end{minipage}%
\begin{minipage}[c]{.5\textwidth}
\includegraphics[width=.7\textwidth,angle=-90]{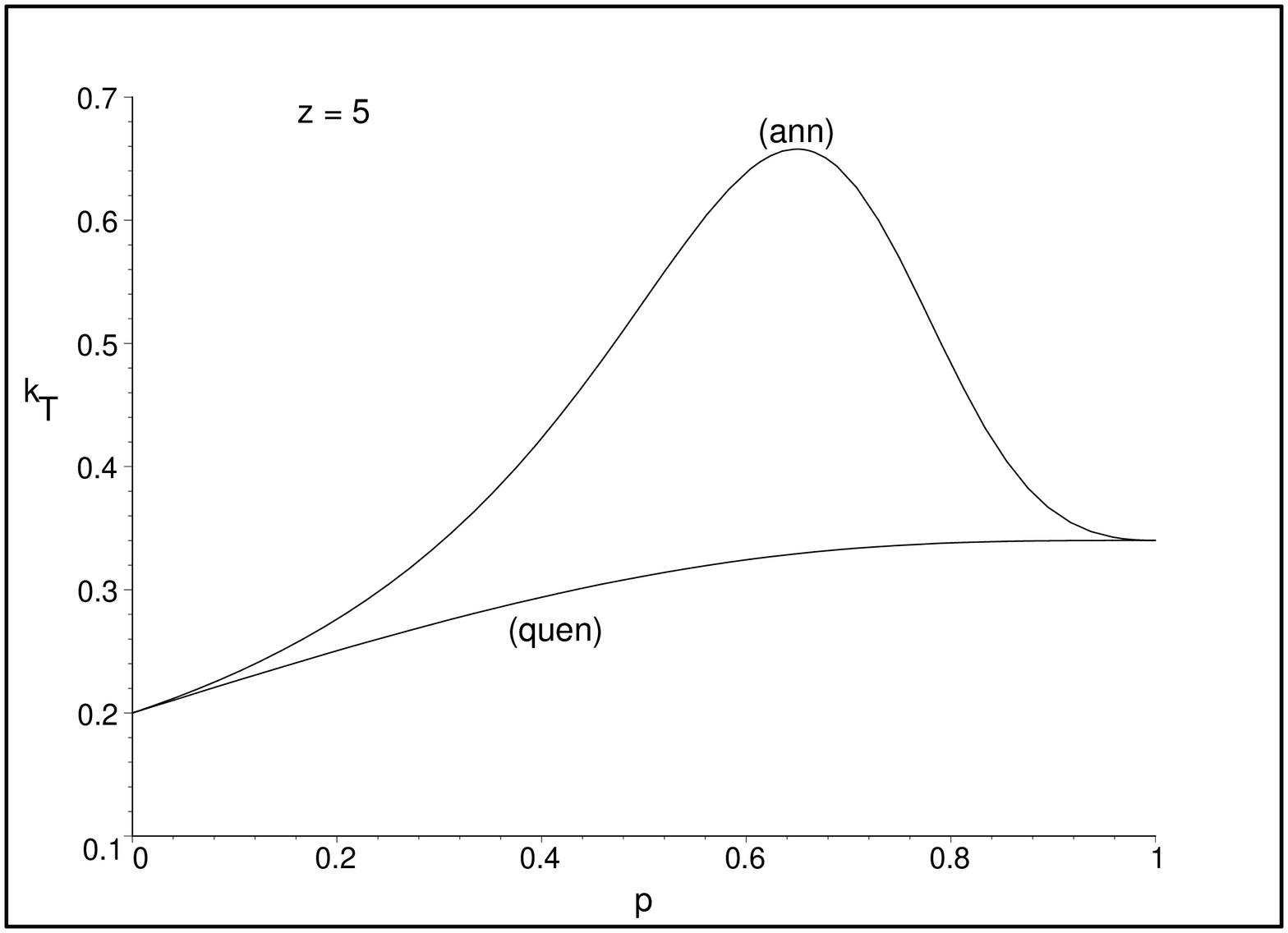}
\end{minipage}
\caption{\label{Fig10} {\small \sl Compressibility $k_T$ as 
a function of $p$ for  
the annealed and for the quenched disorder cases.} }
\end{center}
\end{figure}

Finally, we find that as $z \to \infty$, the mean particle density $n^{(quen)}$ tends to
\begin{equation}
\lim_{z \to \infty} n^{(quen)}(p) = 1 - \frac{p}{1 + p^2},
\end{equation}
which contradicts apparently the behavior observed in the annealed disorder case, where
we found that $\lim_{z \to \infty} n^{(ann)}(p) \equiv 1$, regardless of the value of $p$, (provided that $p < 1$);
it also
differs from our predictions for the case of a regular placement of catalytic sites, for which
$\lim_{z \to \infty} n^{(reg)}(L) = 1 - p$ for $p \leq 1/2$ and  
$\lim_{z \to \infty} n^{(reg)} \equiv 1/2$ for $p = 1/2$ and $p = 1$. Note also (despite of
the fact that here $p = 1$ also appears as a special point in regard to the large-$z$ behavior 
of the disorder-averaged pressure) that here, contrary to the annealed disorder case, 
$n^{(quen)}(p)$  does not show any discontinuity 
in the limit $p \to 1$ at $z = \infty$.

To close this final 
section we discuss
the behavior of the compressibility in the  quenched disorder case. 
From eqs.(\ref{resultat integral}) and (\ref{density3}) we find that in the small-$p$ limit
the compressibility follows
\begin{eqnarray}
\beta^{-1} k_T^{(quen)} ={\frac {1}{z}}+{\frac { \left( 7+6\,z+2\,{z}^{2} \right) z 
}{ \left( 1+3\,z+{z}^{2} \right) ^{2}}}\,p + {\cal O}(p^2),
\end{eqnarray}
while in the opposite limit $p \sim 1$ it is described by a more complicated expression of the form:
\begin{eqnarray}
\beta^{-1} k_T^{(quen)} = k_T^{(reg)}(L=1 \, {\rm or} \, 2) + \frac{3 z \sqrt{1 + 4 z} \left( \sqrt{1 + 4 z}  + 1 
- 2 z \right)}{2 \left(4\,{z}^{3}+9\,{z}^{2}+6\,z+1\right)}\left( 1 - p \right)^{3} + {\cal O}\left(\left( 1 - p \right)^{4}\right)
\end{eqnarray}
It follows then that the coefficient in the term proportional to $(1 - p)^3$ is positive only for $z < 2$ and
 negative for $z > 2$.
This implies, in view of the discussion 
presented at the end of the previous section, that 
for $z < 2$ the compressibility  $k_T^{(quen)}$ is a non-monotonic function of $p$. 
On the other hand, 
for $z > 2$ the compressibility  $k_T^{(quen)}$ seems to be always increasing with $p$.
 This is distinct from
the behavior found in the annealed disorder case when $k_T^{(ann)}$
is non-monotonic function  for any $z$.

In Figs.10 we depict 
the behavior of the compressibility as a function of $p$ 
in the quenched disorder case and compare it to the behavior 
observed in  the annealed disorder case. 
These figures suggest that for $z < 2$ the compressibility
$k_T^{(quen)}$ is a non-monotonic function of $p$, while for $z > 2$, (as exemplified here 
by the case $z = 5$) it is monotonic. 
We also find 
that 
the most dramatic difference between the three different modes of placing
the catalytic sites is seen in $k_T$.

\section{Conclusion.}

In this paper we have studied  
the properties
of the  catalytically-activated annihilation
$A + A \to 0$ reaction 
on  a  one-dimensional lattice, 
in which some lattice sites 
possesses special "catalytic" properties; reaction
takes place when at least one of 
two neighboring adsorbed $A$ particles, undergoing 
continuous exchanges with a particles reservoir, sits on a
catalytic site.

We have focused here 
on three different
situations. First, we have considered 
the case when the catalytic sites are placed periodically, forming
a regular sublattice, in which case 
we obtained
the exact solution in a straightforward manner.
 Next, we turned to the disordered case and studied
the reaction properties for both  annealed and quenched randomness 
in the distribution of the catalytic sites.
We have shown that in the annealed disorder case
 the model reduces to a one-dimensional lattice gas with an
effective three-particle repulsive interaction. We have 
developed a combinatorial formulation
of the model which allowed us to 
obtain the exact solution. Next,
we have demonstrated that in the (most complex) situation with quenched disorder 
the problem 
of computation of the average logarithm of the partition function
can be reduced to the problem of 
the enumeration of all
possible interconnected clusters in finite lattices 
with a fixed number of catalytic
sites. We have calculated the weights of these clusters exactly, using a 
combinatorial procedure, and we have found 
an exact expression
 for the disorder-averaged pressure. 
 
Apart from the results on the disorder-averaged pressure,  we have determined
 exact asymptotic
expressions for the mean-particle density and for the compressibility. We have 
shown that the
behavior of these properties is substantially different in systems with 
 annealed and with quenched
disorder. 
Both differ considerably from the mean-field result of eq.(\ref{no}).
Furthermore, we have observed that 
in systems with annealed disorder  the mean particle density tends to unity in the limit
when the chemical potential
$\mu$ tends to infinity,
and this for any limited catalytic sites density. 
On the other hand, we have established that 
the mean particle density in the quenched disorder case
tends to a finite value, namely to  $(1 - p + p^2)/(1 + p^2) < 1$ when $\mu$
tends to infinity. As well, 
we have demonstrated that in
the $annealed$ disorder case the compressibility appears to be a non-monotonic
function of $p$ for any $\mu$, while in the quenched disorder case the compressibility
shows a non-monotonic behavior as a function 
of $p$ only for $\mu \leq \mu_{crit}
= \beta^{-1} \ln(2)$. 

We close by noting 
that the model studied here 
furnishes another example (see, e.g.,  Refs.~\cite{der1,der2,der3,ranis})
of a 1D Ising-type system 
with random multisite
interactions 
which admits an exact solution. 

\section{Acknowledgment.}

The support of the Deutsche Forschungsgemeinschaft
and of the Fonds der Chemischen Industrie is gratefully
acknowledged.

\pagebreak

\section*{Appendix A}
\addcontentsline{toc}{section}{Appendix A}

\setcounter{section}{1} 
\renewcommand{\theequation}{\Alph{section}.\arabic{equation}}
\setcounter{equation}{0}

The partition function in eq.(\ref{+}) can be represented as
\begin{equation}
\left<Z_N(\zeta)\right> = I_1 - I_2,
\end{equation}
where
\begin{eqnarray}
\label{j1}
I_1 &=& \frac{\displaystyle \Big(z (1-p)\Big)^{N+1}}{2 \pi i} \oint_{{\cal C}}  \frac{d \tau}{\tau} \;
\tau^{- N} \Big(\frac{1-p}{p} + \tau - \tau^2\Big)
\; \times \nonumber\\
&\times& \Big(\frac{z (1-p)^2}{p} - \frac{(1-p)}{p}(1 + z (1-p)) \tau -
\tau^2  + \tau^3 \Big)^{-1},
\end{eqnarray}
and
\begin{eqnarray}
\label{j2}
I_2 &=&  \frac{(1-p)}{2 \pi i p} \oint_{{\cal C}}  d \tau  \;  (1 - \tau)
 \Big(\frac{1 - p - p \tau + (1 - p) \tau^2}{(1-p)
(1-\tau) }\Big)^{N+2}
\; \times \nonumber\\
&\times& \Big(\frac{z (1-p)^2}{p} - \frac{(1-p)}{p}(1 + z (1-p)) \tau -
\tau^2  + \tau^3 \Big)^{-1}.
\end{eqnarray}

To evaluate the integrals in eqs.(\ref{j1}) and (\ref{j2}), 
let us first express the
denominator of the integrands in terms of elementary fractions; this gives
\begin{eqnarray}
\label{denom}
&&\frac{1}{\displaystyle \Big(\frac{z (1-p)^2}{p} - \frac{(1-p)}{p}(1 + z (1-p)) \tau -
\tau^2  + \tau^3 \Big)} = - \frac{1}{t_1 (t_1 - t_2) (t_1 - t_3) (1 - t_1^{-1} \tau)} - \nonumber\\
&-& \frac{1}{t_2 (t_2 - t_1) (t_2 - t_3) (1 - t_2^{-1} \tau)}
-
\frac{1}{t_3 (t_3 - t_2) (t_3 - t_1) (1 - t_3^{-1} \tau)},
\end{eqnarray}
where $t_1$, $t_2$ and $t_3$ are three roots of the cubic equation
\begin{equation}
\label{cubic}
t^3 - t^2  - \frac{(1-p)}{p}(1 + z (1-p)) t + \frac{z (1-p)^2}{p} = 0.
\end{equation}
Expanding next $(1 - t_i^{-1} \tau)$ in a Taylor series with respect to
$\tau$  and taking advantage of the definition of the lattice
delta-function in eq.(\ref{delta}), we find that
$I_1$ of eqs.(\ref{j1}) is given by
\begin{eqnarray}
\label{mmmm}
I_1&=&   \displaystyle  \Big\{
\frac{ \displaystyle p t_1^2 - p t_1 - (1 - p)}{ \displaystyle p  (t_1 - t_2) (t_1 - t_3)}
\Big(\frac{ \displaystyle z (1-p)}{ \displaystyle t_1}\Big)^{N+1} + \nonumber\\
&+& \frac{ \displaystyle p t_2^2 - p t_2 - (1 - p)}{ \displaystyle p (t_2 - t_1) (t_2 - t_3)}
\Big(\frac{ \displaystyle z (1-p)}{ \displaystyle t_2}\Big)^{N+1} + \nonumber\\
&+& \frac{ \displaystyle p t_3^2 - p t_3 - (1 - p)}{ \displaystyle p (t_3 - t_2) (t_3 - t_1)}
\Big(\frac{ \displaystyle z (1-p)}{ \displaystyle t_3}\Big)^{N+1}
\Big\}.
\end{eqnarray}
On the other hand, the function in eq.(\ref{denom}) and also the expression
\begin{equation}
 \Big(\frac{1 - p - p \tau + (1 - p) \tau^2}{(1-p)
(1-\tau) }\Big)^{N+2}
\end{equation}
are both analytic functions of the variable $\tau$. 
Hence, in virtue of eq.(\ref{delta}), it follows that 
the integral in eq.(\ref{j2}) equals  zero and hence, $I_2  = 0$.

Furthermore, 
in terms of the auxiliary functions $R$ and $Q$, 
Eqs.(\ref{Q}) and (\ref{R}),
the roots  of the cubic equation (\ref{cubic}) 
can be written as follows \cite{abr}:
\begin{eqnarray}
\label{roots1}
t_1 &=&
\frac{1}{3}  + \Big[\Big(R + \sqrt{Q^3 + R^2} \Big)^{1/3} + \Big(R -
 \sqrt{Q^3 + R^2} \Big)^{1/3}  \Big],
\end{eqnarray}
\begin{eqnarray}
\label{roots2}
t_2 &=&
\frac{1}{3} - \frac{1}{2} \Big[\Big(R + \sqrt{Q^3 + R^2} \Big)^{1/3} + \Big(R -
 \sqrt{Q^3 + R^2} \Big)^{1/3} \Big] - \nonumber\\
&-& \frac{i \sqrt{3}}{2} \Big[\Big(R + \sqrt{Q^3 + R^2}
 \Big)^{1/3} - \Big(R -
 \sqrt{Q^3 + R^2} \Big)^{1/3} \Big],
\end{eqnarray}
and
\begin{eqnarray}
\label{roots3}
t_3 &=&
\frac{1}{3} - \frac{1}{2} \Big[\Big(R + \sqrt{Q^3 + R^2} \Big)^{1/3} + \Big(R -
 \sqrt{Q^3 + R^2} \Big)^{1/3} \Big] + \nonumber\\
&+& \frac{i \sqrt{3}}{2} \Big[\Big(R + \sqrt{Q^3 + R^2}
 \Big)^{1/3} - \Big(R -
 \sqrt{Q^3 + R^2} \Big)^{1/3} \Big].
\end{eqnarray}
Note next that
the characteristic sum
$Q^3 + R^2$ is less or equal to zero
for any value of the parameters $p$ and $z$; hence,
all three
roots of the cubic equation (\ref{cubic}) are real.
Noticing also that the ratio $ R/\sqrt{-Q^3}$  is bounded,
$-1 \leq R/\sqrt{-Q^3} \leq 1$,  we find that the
 roots can be expressed in a more convenient fashion as:
\begin{eqnarray}
\label{ll1}
t_1 = \frac{1}{3} + 2 \sqrt{-Q} \cos\Big(\frac{1}{3}
\arccos\Big(\frac{R}{\sqrt{-Q^3}}\Big)\Big),
\end{eqnarray}
\begin{eqnarray}
\label{ll2}
t_2 = \frac{1}{3} -  2 \sqrt{-Q} \sin\Big(\frac{1}{3}
\arcsin\Big(\frac{R}{\sqrt{-Q^3}}\Big)\Big),
\end{eqnarray}
and
\begin{eqnarray}
\label{ll3}
t_3 = \frac{1}{3}  -  2 \sqrt{-Q} \sin\Big(\frac{\pi}{6} + \frac{1}{3}
\arccos\Big(\frac{R}{\sqrt{-Q^3}}\Big)\Big).
\end{eqnarray}
Noticing now that all roots $t_i$, $(i=1,2,3)$, of eq.(\ref{cubic}) obey
\begin{equation}
p t_i^2 - p t_i - (1 - p) = 2 (1 - p) (t_i - 1) \Big(\frac{z (1-p)}{2 t_i}\Big),
\end{equation}
while
\begin{equation}
t_1 - t_2 = 2 \sqrt{- 3 Q} \cos\Big(\frac{\pi}{6} + \frac{1}{3}
\arccos\Big(\frac{R}{\sqrt{-Q^3}}\Big)\Big),
\end{equation}
\begin{equation}
t_1 - t_3 = 2 \sqrt{- 3 Q} \cos\Big(\frac{\pi}{6} - \frac{1}{3}
\arccos\Big(\frac{R}{\sqrt{-Q^3}}\Big)\Big),
\end{equation}
and
\begin{equation}
t_2 - t_3 = 2 \sqrt{- 3 Q} \sin\Big(\frac{1}{3}
\arccos\Big(\frac{R}{\sqrt{-Q^3}}\Big)\Big),
\end{equation}
and substituting the results in eqs.(\ref{ll1}) to (\ref{ll3})
into eq.(\ref{mmmm}),
we find, eventually, the result in eq.(\ref{ZZ}).

\section*{Appendix B}
\addcontentsline{toc}{section}{Appendix B}

\setcounter{section}{2} 
\renewcommand{\theequation}{\Alph{section}.\arabic{equation}}
\setcounter{equation}{0}

Summing 
${\cal N}_K^{(n)}(\{l_{n}\}|N)$
over all the interval realizations $\{l_{n}\}$
obeying the conservation law in eq.(\ref{dina}), 
 and using the integral representation of the 
Kronecker function in eq.(\ref{delta}),
we obtain, after some straightforward calculations, that:
\begin{equation}
\sum_{\{l_{n}\}} J^{(S)}_{(n)}(\{l_{n}\}|K|N)
= 2\;{K-1-n \choose n-1}\;{N-K-1 \choose N_{nc}-n-1},
\end{equation}
while
\begin{equation}
\sum_{\{l_{n}\}} J^{(B)}_{(n)}(\{l_{n}\}|K|N)
= (N_{nc}-n)\;{K-2-n \choose n-1}\;{N-K-1 \choose N_{nc}-n-2}.
\end{equation}
Performing next the summation over
all possible numbers $n$ of 
subintervals 
in a $K$-cluster,  we
find that the total number $N_K(N_{nc}|N)$ of $K$-clusters 
for all possible realizations of an $N$-chain containing a fixed number
$N_{nc}$ of non-catalytic sites is given by
\begin{eqnarray}
N_K(N_{nc}|N)&=&
\left(1-\delta(K,N)\right)\;\left(1-\delta(K,1)\right)\;  
\Big\{2\sum_{n=1}^{[K/2]}\;{K-1-n \choose n-1}\;{N-K-1
\choose
N_{nc}-n-1}\; + \nonumber\\
&+& \;\sum_{n=1}^{[(K-1)/2]}\;(N_{nc}-n)\;{K-2-n \choose
n-1}\;{N-K-1 \choose N_{N_{nc}}-n-2}\Big\} + \nonumber\\
&+&\;\delta(K,N)\;\Big\{2\sum_{n=1}^{[K/2]}\;\delta(n,N_{nc})\;{K-1-n
\choose
n-1}\;+\nonumber\\
&+&\;\;\sum_{n=1}^{[(K-1)/2]}\;\delta(n,N_{nc}-1)\;{K-2-n
\choose
n-1}\;+
\;\sum_{n=1}^{[(K+1)/2]}\;\delta(n,N_{nc}+1)\;{K-n
\choose
n-1}\;\Big\}+\nonumber\\
&+&\;\delta(K,1)\;(1-\delta(N_{nc},1))\;(1-\delta(N_{nc},0))\;N_{nc}\;{N-2
\choose N_{nc}-2}.
\end{eqnarray}
Now, representing the weights $w_{K,N}(p)$ 
of $K$-clusters as
\begin{equation}
 w_{K,N}(p)=w_{K,N}^{(B)}(p)+w_{K,N}^{(S)}(p),
\end{equation}
where $w_{K,N}^{(B)}(p)$ and $w_{K,N}^{(S)}(p)$ 
denote the weights of the "bulk" and "surface" $K$-clusters,
 respectively, we find,
summing over all possible 
numbers $N_{nc}$ of non catalytic sites, that these weights are defined explicitly by
\begin{equation}
\label{bulk}
w_{K,N}^{(B)}(p)=p^N \sum_{n=1}^{[(K-1)/2]}{K-2-n \choose n-1}
\sum_{N_{nc}=0}^\infty
(N_{nc}-n)\;\left(\frac{1-p}{p}\right)^{N_{nc}}{N-K-1 \choose
N_{nc} -n -2}
\end{equation}
and
\begin{equation}
\label{surface}
w_{K,N}^{(S)}(p) = 2 p^N \sum_{n=1}^{[(K-1)/2]}{K-1-n \choose n-1}
\sum_{N_{nc}=0}^\infty \; \left(\frac{1-p}{p}\right)^{N_{nc}}{N-K-1
\choose N_{nc} -n -1}.
\end{equation}
Noticing next that
\begin{eqnarray}
\sum_{N_{nc}=0}^{\infty}
(N_{nc}-n)\;\left(\frac{1-p}{p}\right)^{N_{nc}} {N-K-1 \choose
N_{nc} -n -2}
=\frac{(1-p)^{n+2}}{p^{n+N-K+1}} \;\left[(1-p)(N-K-1)+2\right],
\end{eqnarray}
we obtain that $w_{K,N}^{(B)}(p)$ obeys
\begin{equation}
w_{K,N}^{(B)}(p)=(1 - p)^2 p^{K - 1} \sum_{n=1}^{[(K-1)/2]} \left[(1-p)(N-K-1)+2\right] {K-2-n \choose
n-1}\;\left(\frac{(1-p)}{p}\right)^n,
\end{equation}
and consequently, in virtue of the definition of the Fibonacci polynomials, eq.(\ref{fibo}),
the total weight of the "bulk" $K$-clusters can be expressed by
\begin{eqnarray}
w_{K,N}^{(B)}(p)&=&p^{(K-1)/2}\;(1-p)^{(K+3)/2}\; \times \nonumber\\
&\times& \left[(1-p)(N-K-1)+2\right]
\; F_{K-2}\left(\sqrt{\frac{p}{1-p}}\right).
\end{eqnarray}
In similar fashion, we find that the total weight of "surface" $K$-clusters 
$w_{K,N}^{(S)}(p)$ obeys
\begin{equation}
w_{K,N}^{(S)}(p) = 2p^{K/2}\;(1-p)^{(K+2)/2}\;
F_{K-1}\left(\sqrt{\frac{p}{1-p}}\right).
\end{equation}
Combining these results, we arrive eventually at the general 
formulae in eqs.(\ref{w}) and (\ref{ww}).

\section*{Appendix C}
\addcontentsline{toc}{section}{Appendix C}

\setcounter{section}{3} 
\renewcommand{\theequation}{\Alph{section}.\arabic{equation}}
\setcounter{equation}{0}

In order to evaluate the limiting behavior of the rather complicated 
sums entering eq.(\ref{calcul}), it is expedient to 
introduce an auxiliary generating function of the form:
\begin{equation}\label{fonction generatrice}
{\cal F}(\xi,\tau)=\sum_{N=3}^\infty {\cal G}_N(\xi)  \tau^N,
\end{equation}
where
\begin{equation}
\label{GN}
{\cal G}_N(\xi) = \sum_{K=2}^{N-1} \frac{w_{K,N}(p)}{N} \xi^K.
\end{equation}
Once ${\cal F}(\xi,\tau)$ and ${\cal G}_N(\xi)$ are determined, one  
obtains $\beta P^{(quen)}_3(p)$ directly. As one may verify readily,  $\beta P^{(quen)}_3(p)$
in eq.(\ref{calcul}) can be expressed in terms of ${\cal G}_N(\xi) $ as
\begin{eqnarray}
\label{gs}
\beta P^{(quen)}_3(p) &=& {\cal G}_{\infty}(\xi = 1) \ln\Big(\frac{z^2 \tau_1^2}{\sqrt{1 + 4 z}}\Big)\;-\;
\left.\frac{\partial {\cal G}_{\infty}(\xi)}{\partial \xi}\right|_{\xi = 1} \ln{\tau_2} -\nonumber\\
&-& \sum_{n = 1}
\frac{(\tau_2/\tau_1)^{2 n}}{n} {\cal
G}_{\infty}\left(\xi = (-\frac{\tau_2}{\tau_1})^{n}\right).
\end{eqnarray}
We turn now to the calculation of the generating function 
in eq.(\ref{fonction generatrice}).
Substituting 
the explicit form of $w_{K,N}(p)$, eq.(\ref{w}), 
into eq.(\ref{GN}), and this in eq.(\ref{fonction generatrice}), and 
interchanging then in the final result
the order of summations (over $K$ and $N$), 
we find that ${\cal F}(\xi,\tau)$ can be 
represented as a sum of 
two components,
\begin{equation}\label{F}
{\cal F}(\xi,\tau) = F_1(\xi,\tau)+F_2(\xi,\tau),
\end{equation}
where
\begin{eqnarray}\label{FB}
F_1(\xi,\tau)&\equiv&\frac{(1 - p)^{5/2}}{p^{1/2}}\sum_{K=2}^\infty \Big(p (1-p) \xi^2\Big)^{\frac{K}{2}}
F_{K-2}\left(\sqrt{\frac{p}{1-p}}\right)
\Big(\sum_{N=K+1}^\infty\frac{N - K - 1}{N} \tau^N\Big),\nonumber\\
\end{eqnarray}
and
\begin{eqnarray}
F_2(\xi,\tau)&=& 2 \frac{(1 - p)^{3/2}}{p^{1/2}}   \sum_{K=2}^\infty \Big(p (1-p) \xi^2\Big)^{\frac{K}{2}} F_{K}\left(\sqrt{\frac{p}{1-p}}\right)
 \Big(\sum_{N=K+1}^\infty\frac{\tau^N}{N}\Big).
\end{eqnarray}
Using next the evident integral equality
\begin{equation}
\frac{1}{N} = \int_{0}^{1} dv \; v^{N - 1},
\end{equation}
as well as the explicit expression of the generating function of the Fibonacci polynomials \cite{fibo}
\begin{equation}
\sum_{K = 1}^{\infty} F_K(x) \tau^K = \frac{\tau}{\tau^2 + x \tau - 1}
\end{equation}
we obtain the following integral
representations of $F_1(\xi,\tau)$ and $F_2(\xi,\tau)$:
\begin{eqnarray}\label{representation integrale de F1}
F_1(\xi,\tau)&=&(1-p)^4p\tau^5\xi^3\;\int_0^1\frac{v^4 {\ud v} }{(1-\tau v)^2}\frac{1}{\Big(1-pv\tau\xi
-p(1-p)(v\tau\xi )^2\Big)},
\end{eqnarray}
and
\begin{eqnarray}\label{representation integrale de F2}
F_2(\xi,\tau)&=&2(1-p)^3p^{3/2}\tau^3\xi\;\int_0^1\frac{v^2 {\ud v}}{(1- \tau v)}\frac{1-p+v\tau\xi}{\Big(1-pv\tau\xi
-p(1-p)(v\tau\xi )^2\Big)}.
\end{eqnarray}
Now, the asymptotic behavior of 
behavior of ${\cal G}_N(\xi)$ as $N\to\infty$ 
can be deduced, in a standard fashion, 
by analysing the critical behavior
of the generating function ${\cal F}(\xi ,\tau)$ 
in the vicinity of the singularity closest to the
origin  \cite{wilf}; that is, here, 
in the vicinity of $\tau=1$. Expanding 
${\cal F}(\xi,\tau)$ 
in the vicinity of the 
singular point  $\tau=1$, we obtain
\begin{eqnarray}
{\cal F}(\xi,\tau)&=&\frac{(1-p)^4p\xi^3}{1-p\xi -p(1-p)\xi
^2}\;\frac{1}{1-\tau}\;+\nonumber\\
&+&\; \frac{2(1-p)^3p^{3/2}\tau^3\xi (1-p+\xi)}{1-p\xi
-p(1-p)\xi ^2}\ln(1-\tau)\;+\;o\left(\ln(1-\tau)\right).
\end{eqnarray}
Next, in virtue of the 
Tauberian theorems \cite{wilf}, 
it follows that 
${\cal G}_N(\xi)$ exhibits 
the following asymptotical behavior
as $N\to\infty$:
\begin{equation}
\label{inf}
{\cal G}_N(\xi)\sim\frac{p(1-p)^4\xi^3}{1-p\xi-p(1-p)\xi^2} -  \left( \frac{2(1-p)^3p^{3/2}\tau^3\xi (1-p+\xi)}{1-p\xi
-p(1-p)\xi ^2}\right) \frac{1}{N} \;+\; o\left(\frac{1}{N}\right),
\end{equation}
and consequently, we find from eq.(\ref{inf}) that the particular values of the function ${\cal G}_N(\xi)$
entering eq.(\ref{gs}) are given explicitly by:
\begin{eqnarray}
\label{ggs}
&&{\cal G}_{\infty}(\xi=1)=p(1-p)^2; \;\;\; 
\left. \frac{\partial {\cal G}_{\infty}(\xi)}{\partial \xi}\right|_{\xi = 1} = p \; (p^2-3p+3), \nonumber\\
\end{eqnarray}
and
\begin{eqnarray}
\label{gggs}
&&{\cal G}_{\infty}\left(\xi=\left(-\frac{\tau_2}{\tau_1}\right)^n\right)=
\frac{p(1-p)^4 \left(-\tau_2/\tau_1\right)^{3n}}{1-p(-1)^n \left(\tau_2/\tau_1\right)^n-p(1-p)\left(\tau_2/\tau_1\right)^{2n}}.
\end{eqnarray}
Substituting the results in eqs.(\ref{ggs}) and (\ref{gggs}) 
into eq.(\ref{gs}), we thus find that $\beta P^{(quen)}_3(p)$ obeys
\begin{eqnarray}\label{resultat33}
\beta P^{(quen)}_3(p) &=& p \Big(5 - 7 p + 3 p^2\Big) \ln\left(\frac{\sqrt{1 + 4 z} + 1}{2}\right)
- \frac{p (1 - p)^2}{2} \ln(1 + 4 z) + \nonumber\\
 &+& p (1-p)^4 \; \sum_{n=1}^\infty \frac{(-1)^n}{n} 
\displaystyle \frac{\left(\tau_2/\tau_1\right)^{5 n}}{1-p (-1)^n \left(\tau_2/\tau_1\right)^n-p(1-p)
 \left(\tau_2/\tau_1\right)^{2n}},
\end{eqnarray}
which, in combination with the expression for $\beta P_1^{(quen)}(p)$, eq.(\ref{single}), leads to 
eq.(\ref{resultat}).

\end{document}